\documentclass[12pt,letterpaper]{article}
\usepackage{latexsym, amsmath, amssymb, amsthm, enumerate, tikz, graphicx, wrapfig, multirow, comment, bbm, listings,subcaption, authblk, import,hyperref,csquotes, appendix,booktabs,abstract}

\usepackage[nodisplayskipstretch]{setspace}
\usepackage{booktabs,caption,cellspace}
\usepackage{algorithm}
\usepackage{algpseudocode}
\usepackage{mathtools}
\numberwithin{equation}{section}

\theoremstyle{remark}

\theoremstyle{definition}

\usepackage{lineno}

\usepackage{tikz}
\usetikzlibrary{bayesnet}
\tikzstyle{doublecircle} = [latent, double, double distance=2pt, outer sep=1pt]
\usetikzlibrary{shapes,decorations,arrows,calc,arrows.meta,fit,positioning}
\tikzset{
	-latex,auto
	state/.style ={ellipse, draw, minimum width = 1cm},
  el/.style = {inner sep=3pt, align=left, sloped},
  point/.style = {circle, 
                  draw = black, 
                  inner sep = 0.04cm,
                  fill = black,
                  node contents={}},
  square/.style={rectangle, 
                 draw = black, 
                 inner sep = 0.08cm,
                 node distance = 0.8cm and 0.8cm,
                 fill=black,  
                 node contents={}},
  bidirected/.style={Latex-Latex,dashed}
}

\usepackage[
natbib=true,
style=nature,
uniquename=false,
hyperref=true,
]{biblatex}

\DeclareSortingTemplate{bibsort}{
  \sort{
    \field{author}
    \field{editor}
    \field{translator}
  }
  \sort{
    \field{title}
  }
}

\title{US COVID-19 school closure was not cost-effective, but other measures were.}

\author{
Nicholas J. Irons\thanks{
Department of Statistics and Leverhulme Centre for Demographic Science, University of Oxford. Corresponding author. Email: 
 \href{mailto:nicholas.irons@stats.ox.ac.uk}{nicholas.irons@stats.ox.ac.uk}.},
Adrian E. Raftery\thanks{
Departments of Statistics and Sociology, University of Washington.}. }

\date{\today}

\begin{document}

\maketitle


\begin{abstract}
Non-pharmaceutical interventions (NPIs) in response to the COVID-19 pandemic
necessitated a trade-off between the health impacts of viral spread and the social and economic costs of restrictions \citep{chetty-impacts,betthauser-learning-review,psacharopoulos-learning,ferguson-npi-impacts}. 
We conduct a cost-effectiveness analysis of NPI policies enacted at the state level in the United States in 2020. 
Although school closures reduced viral transmission, their social impact in terms of student learning loss was too costly, depriving the nation of \$2 trillion (USD2020), conservatively, in future GDP \citep{psacharopoulos-learning,hanushek-learning,betthauser-learning-review}.
Moreover, this marginal trade-off between school closure and COVID deaths was not inescapable: a combination of other measures would have been enough to maintain similar or lower mortality rates without incurring such profound learning loss. 
Optimal policies involve consistent implementation of mask mandates, public test availability, contact tracing, social distancing orders, and 
reactive workplace closures, with no closure of schools beyond the usual 16 weeks of break per year. 
Their use would have reduced the gross impact of the pandemic in the US in 2020 from \$4.6 trillion to \$1.9 trillion
and, with high probability, saved over 100,000 lives.
Our results also highlight the need to address the substantial global learning deficit incurred during the pandemic \citep{unicef-education-crisis,betthauser-learning-review}.
\end{abstract}

\section{Introduction}
\label{sec:intro}

In the year prior to the arrival of COVID vaccines and other pharmaceutical interventions, non-pharmaceutical interventions (NPIs)---including school and workplace closures, social distancing, masking, testing, and contact tracing---were the primary tools for mitigating the spread of SARS-CoV-2. 
The use of NPIs posed significant challenges to decision-makers at every level of government, who were forced to make difficult and consequential real-time decisions with limited data and amidst contentious political debate \citep{adolph-politics-timing}. 
While they substantially reduced viral transmission, extended lockdowns had severe deleterious social and economic consequences globally---including disrupted economic output, job loss, and student learning loss \citep{chetty-impacts,unicef-education-crisis}---on top of the already staggering health impacts of the pandemic. These impacts---health, economic, and social---were felt disproportionately by marginalized populations \citep{chetty-impacts,tai-covid-minorities,betthauser-learning-review}.

To address the NPI planning problem in a principled way, we develop a statistical decision framework and conduct a cost-effectiveness analysis of non-pharmaceutical intervention policies enacted at the state level in the United States in 2020.
Our analysis is composed of three steps. We first build a Bayesian epidemiological model estimating SARS-CoV-2 prevalence and transmission rates in each state over time based on prior work leveraging random sample testing surveys to debias clinical COVID data \citep{irons}. We next estimate the effects of NPIs on viral transmission in all states jointly using a Bayesian hierarchical regression model controlling for temporal autocorrelation and endogenous behavioral responses linked to fear of infection. Finally, we couple these estimates with monetary costs associated with the social, economic, and health consequences of infection and NPIs drawn from the literature in order to quantitatively evaluate the efficacy and gross impacts of the policy schedules implemented during the pandemic and to derive strategies that optimally navigate the trade-off between restrictions and viral spread. 

The literature studying non-pharmaceutical interventions in response to COVID-19 and past pandemics is vast.
\citet{brodeur-review,bloom-macro-review} provide reviews of the economics literature. 
Our work is most closely related to 
studies: estimating associations and inferring causal effects of NPIs on viral transmission \citep{flaxman-effects,brauner-effects,sharma-effects,chernozhukov-masks}; quantifying the gross health and economic impacts of pandemics and the associated policy response \citep{thunstrom-costs-benefits,greenstone-distancing,hall-vsl,summers,kaplan-covid-political-economy,prager-flu-cost,ferguson-npi-impacts}; and modeling the (optimal) control of epidemics and the cost-effectiveness of non-pharmaceutical interventions appearing in the economics
\citep{farboodi2021,eichenbaum-macro,acemoglu-targeted,adda-oc,alvarez-oc,barrot-business-effect}
and public health literature
\citep{ferguson-strategies,halloran-npi-containment,xue-school-cost,halder-flu-cost,perlroth-flu-costs,keogh-brown-covid-cost, brown-flu-school-cost}.
Throughout the text, we highlight how the results of our models compare to these studies.

We build upon this body of work to address gaps limiting its value in informing policy. 
Firstly, we study 
the optimal control of a pandemic using statistical decision theory.
We take a data-driven approach to the NPI decision process, estimating and accounting for uncertainty in key parameters, including viral prevalence, reproduction numbers, and the effects of NPIs and other endogenous and exogenous factors on transmission rates. In particular, we produce probabilistic estimates of SARS-CoV-2 prevalence over time, which are necessary to properly account for the magnitude and uncertainty of costs associated with infections.   
Our model is able to capture the complex and stochastic temporal trends of SARS-CoV-2 transmission (e.g., multiple waves, super-spreader events, the introduction of new infections via travel, and random fluctuations) which can be missed by standard deterministic epidemiological models.
This allows us to define and evaluate realistic counterfactual scenarios under different NPI policies conditional on what was observed during the pandemic.
Given the largely unpredictable nature of SARS-CoV-2 transmission, we find that the structure of the optimal NPI strategy is remarkably simple and consistent across time and space, with the planner required to respond to COVID dynamics in real time to a minimal degree. 

Secondly, we model the costs and effects on viral transmission of multiple specific NPIs.  
Some previous studies have considered a limited toolkit, focusing on a minimal collection of interventions, such as a single catch-all ``social distancing", ``containment", or ``lockdown" policy \citep{brown-flu-school-cost,xue-school-cost,barrot-business-effect,farboodi2021,acemoglu-targeted,eichenbaum-macro,alvarez-oc}.
This can fail to identify the most effective policies, as we generally have a range of tools at our disposal, and NPIs are known to be more effective in combination 
\citep{perlroth-flu-costs,halder-flu-cost,ferguson-strategies,juneau-review}.
Furthermore, if we consider only a single instrument, we may erroneously conclude that its implementation is cost-effective because we implicitly assume that other policies
are not available.
As we discuss in Section \ref{sec:results-icer}, the cost-effectiveness of any single intervention is context-specific and depends on the other policy options.
We consider a comprehensive set of 11 non-pharmaceutical interventions.
As such, our findings refine the broad qualitative guidance drawn from prior studies in the context of COVID-19---e.g., that ``lockdown'' is cost-effective and optimal when implemented early and stringently \citep{alvarez-oc}. 
By evaluating a range of NPIs, we can disaggregate policies to conclude that testing, tracing, masking, reactive workplace closure, and social distancing measures (not including extended school closure) combine to form an optimal cost-effective strategy.

Thirdly, unlike many studies assessing the economic impact of school closure during pandemics, we factor in costs associated with student learning loss \citep{keogh-brown-covid-cost,halder-flu-cost,brown-flu-school-cost,perlroth-flu-costs}.  
Many studies quantify the total cost of school closure as a sum of direct costs arising from lost productivity of school staff and workplace absenteeism of parents or childcare costs resulting from students staying home. However, the indirect costs of school closure are substantial. Students suffering acute learning loss go on to become less skilled and less productive members of the workforce, which in turn leads to future losses in personal income and national GDP \citep{hanushek-learning}. We account for the net present value of these future losses to society, which can be very large, based on estimates of the cost of learning loss from the education economics literature \citep{hanushek-learning,psacharopoulos-learning,azevedo-school-impacts} and recent estimates of the amount of learning loss accrued during COVID-19 school closures \citep{betthauser-learning-review}.
Considering other indirect costs, we note that: school disruptions and decreases in educational attainment may be associated with various negative health outcomes among students, including depression, anxiety, and decreased life expectancy \citep{viner-school-health,christakis-education-life-loss};\footnote{
Similarly, there is some evidence that COVID-19 restrictions are associated with increases in drug overdose fatalities \citep{wolf-npis-overdose}. Nevertheless, as with school closures (and for the same reasons), we do not take into account potential physical and mental health costs related to other NPI policies.
We discuss this further in Section \ref{sec:discussion}.
} and school closures can cause significant healthcare worker absenteeism, potentially negating some or all of the mortality benefits from school-closure-related reductions in SARS-CoV-2 transmission \citep{lempel-school-cost,bayham-school-workforce}. 
Regarding the former, we do not account for potential downstream physical and mental health costs of school closures as comprehensive causal links and quantitative estimates have not been established. Regarding the latter, we do not account for health impacts related to healthcare personnel absenteeism as quantifying their cost is challenging.  
For these and other reasons, which we discuss in more detail in Section \ref{sec:cost-school}, we believe that our accounting of the costs of school closure is conservative.

In our review of the literature, we found only two cost-benefit analyses of school closure that account for learning loss \citep{xue-school-cost,adda-oc}. Studying pandemic flu, \citet{xue-school-cost} find that school closures are not cost-effective for mild strains 
(such as the 2009 H1N1 virus), but they
generate net benefits 
in the context of more severe pandemics, such as the 1918 Spanish flu.
Similarly, studying historical outbreaks of influenza, gastroenteritis, and chickenpox in France, \citet{adda-oc} determines that school closures were not cost-effective, but that they would become beneficial for slightly more lethal epidemics. 
Notably, \citet{xue-school-cost} model school closure in isolation, i.e., they do not consider the availability of other interventions, and \citet{adda-oc} considers only school and public transport closure. 
Their results concur with a number of other studies (not accounting for learning loss) finding that extended school closures are cost-effective for severe pandemics \citep{dauelsberg-flu-schools,perlroth-flu-costs,milne-flu-cost,kelso-flu-cost,deb-effects}. 
To the contrary, we demonstrate that COVID-19 school closures were not cost-effective.\footnote{
While we find that extended school closure is not cost-effective,
a relevant (and potentially cost-effective) counterfactual
would have been a reactive school closure of limited duration at the beginning of the pandemic (i.e., spring of 2020) compensated by an extended school year stretching into the summer. Such a strategy would not have incurred student learning loss; it would have merely shifted the summer break toward spring to allow for an urgent response to the initial outbreak.
}


Additional methodological contributions of our approach include a novel zero-inflated negative binomial model that flexibly captures well-known reporting idiosyncrasies and over-dispersion in clinical COVID data. As a result, our method eliminates the need for \emph{ad hoc} data cleaning and smoothing procedures that can complicate the analysis pipeline, yield poorly calibrated prediction intervals, and potentially bias transmission rate estimates based on over-smoothed data. Furthermore, we implement a two-stage modeling procedure that first estimates the time-varying effective reproduction number in each US state individually, followed by a joint hierarchical model across states that estimates pooled effects of NPIs on transmission dynamics.
This approach 
allows for efficient Bayesian computation by parallelizing model fits across states. 
While our results are specific to the COVID-19 pandemic, our methods can be used more widely to evaluate public health interventions against infectious disease.

\paragraph{Outline of the text.}
Section \ref{sec:methods} details our methodology, including specifics of the data and implementation, the construction of our models, and the elicitation of costs associated to infections and NPIs. Section \ref{sec:results} reports the baseline results of our models and compares our results to others in the literature. Section \ref{sec:discussion} provides concluding remarks
and discusses qualifications and limitations of our methodology.
The Appendix contains the results of our sensitivity analysis and discusses implications of our methodology and results for the use of incremental cost-effectiveness ratios (ICERs) in infectious disease.

\section{Methods}
\label{sec:methods}

\subsection{Data and implementation}
\label{sec:data}

\paragraph{Data availability.}
All data used are publicly available.
We obtained U.S. state-level daily counts of confirmed COVID cases and deaths in 2020 from the COVID-19 Data Repository by the Center for Systems Science and Engineering (CSSE) at Johns Hopkins University (JHU) \citep{jhu-csse-data}. If a negative number of deaths or cases were reported on a given day---often due to retroactive changes in the reported cumulative death or case count for record deduplication or changes in data reporting by the state government---we assume that the cumulative death (case) count on that day was the correct one and set the number of deaths (cases) incident on prior days to zero until the overall cumulative count is non-decreasing.
We begin modeling viral transmission in each state 
21 days prior to the first day on which more than one death is reported.

We obtained state-level government NPI policies reported daily from the Oxford COVID-19 Government Response Tracker (OxCGRT) \citep{oxcgrt}. In converting the ordinal policy levels to numerical values, we followed OxCGRT's methodology for calculating indices, in which ordinal levels are equally spaced numerically and a targeted (as opposed to general) intervention is treated as a half-step between ordinal levels. We rescale each policy value to lie between 0 and 1, with 1 denoting the most stringent policy. If a policy is not recorded on a given day, we set its value to that on the previous day on which the policy was recorded, or we set it to zero if at the beginning of the study period. We average daily policy values at the weekly level for our NPI regression model.

We obtained state-level counts of SARS-CoV-2 PCR tests administered on each day from the COVID Tracking Project \citep{ctp}. We obtained daily average surface temperature data for the largest city in each state using the Meteostat Python package \citep{meteostat}.

\paragraph{Code availability.}
Data cleaning was conducted in R and Python. All models were fit in R using the CmdStanR package, with MCMC convergence assessed using the diagnostics provided therein \citep{r,cmdstanr}. We used the optimParallel R package for NPI policy optimization \citep{optim-parallel}. To determine the optimal NPI strategy in each state based on the cost function outlined in Section \ref{sec:methods-cost}, we used a combination of 8 random and hand-specified parameter initializations and kept the policy yielding the smallest value of the cost function. In practice, we find that the results of the optimization are robust to the initial parameters, which is reflected in our results. Code to reproduce our analysis will be made available at \url{https://github.com/njirons/covidOC}. 



\subsection{Bayesian epidemiological model}
\label{sec:epi}

We begin by describing the Bayesian epidemiological model used to estimate SARS-CoV-2 prevalence and transmission rates in each US state in 2020. In our two-stage estimation procedure, we first fit this epidemiological model to each state separately. Next, the time-varying basic reproduction numbers $R_0(t)$ output by the model in each state are fed into the Bayesian hierarchical regression model described below in Section \ref{sec:bhm} in which we jointly model transmission rates in all states as a function of NPIs. 


\subsubsection{SEIRD model}

\begin{figure}[!tbp]
\begin{subfigure}[t]{0.5\linewidth}
\centering
  \tikzstyle{every edge}=[draw,>=stealth',semithick]
    \begin{tikzpicture}[->,>=stealth',shorten >=1pt,auto,node distance=1.2cm,scale=1.0,transform shape]
        \node (1) at (-3,0) {$S$};
        \node (2) at (-1.5,0) {$E$};  
        \node (3) at (0,0) {$I$};
        \node (4) at (1.5,1.5) {$R_S$};
        \node (5) at (1.5,-1.5) {$R_D$};
        \node (6) at (3,-1.5) {$D$};
        
        \path (1) edge (2);
        \node at (-2.25,0.4) {$\beta I$};
        \path (2) edge (3);
        \node at (-0.75,0.4) {$\delta$};
        \path (3) edge (4);
        \node at (1.35,0.5) {$\gamma(1-\iota)$};  
        \path (3) edge (5);
        \node at (1.00,-0.5) {$\gamma\cdot\iota$};      
        \path (5) edge (6);
        \node at (2.25,-1.25) {$\mu$};
    \end{tikzpicture}
    \caption{SEIRD model.} 
    \label{fig:seird}
\end{subfigure}
\begin{subfigure}[t]{0.5\linewidth}
\centering
  \tikzstyle{every edge}=[draw,>=stealth',semithick]
    \begin{tikzpicture}[->,>=stealth',shorten >=1pt,auto,node distance=1.2cm,scale=1.0,transform shape]
        \node (1) at (-1.0,0) {$(\tilde{I},\tilde{R},\tilde{D})(w-1)$};        
        \node (2) at (0.0,-1.5) {$u(w)$};
        \node (3) at (2.0,1.5) {$\varepsilon(1:(w-1))$};
        \node (4) at (4.5,-1.5) {$R_0(w)$};
        \node (5) at (4.5,1.5) {$\varepsilon(w)$};
        
        \path (1) edge (2);
        \path (1) edge (4);
        \path (2) edge (4);
        \path (3) edge (4);
        \path (3) edge (1);
        \path (5) edge (4);
        \path (3) edge (2);
    \end{tikzpicture}
    \caption{NPI regression model DAG.}
    \label{fig:dag}
\end{subfigure}
\caption{Graphical models.}
\end{figure}
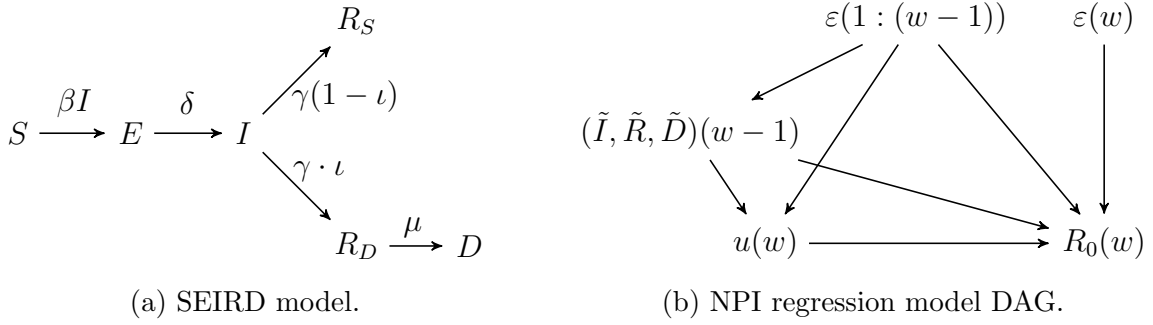

Our discrete-time Bayesian Susceptible-Exposed-Infectious-Removed-Deceased (SEIRD) model builds on the model of \citet{irons}, which was used to estimate state-level SARS-CoV-2 prevalence in the first year of the pandemic based on reported cases, deaths, tests, and random testing surveys.

For a given US state, let $S(t)$ denote the proportion of susceptible people in the state on day $t$, $E(t)$ the proportion exposed but not yet infectious, $I(t)$ the proportion infectious, $R_S(t)$ the proportion recovered (survivors no longer infectious), $R_D(t)$ the proportion no longer infectious who will eventually succumb to the disease, and $D(t)$ the proportion decedent. 
These quantities evolve in time according to the equations
\begin{linenomath*}
\begin{equation}
\begin{cases}
S(t+1) - S(t) &= - \beta(t)S(t)I(t) \\
E(t+1) - E(t) &= \beta(t)S(t)I(t) - \delta E(t) \\
I(t+1) - I(t) &= \delta E(t) - \gamma I(t) \\
R_S(t+1) - R_S(t) &= \gamma(1-\iota)I(t) \\
R_D(t+1)-R_D(t) &= \gamma\cdot\iota I(t) -\mu R_D(t) \\
D(t+1)-D(t) &= \mu R_D(t).
\end{cases}
\label{eq:seird}
\end{equation}
\end{linenomath*}
A graphical model of this process is depicted in Figure \ref{fig:seird}. 

Members of the population move from susceptible to exposed after contact with an infectious person with rate $\beta(t)$, which is allowed to vary in time to account for variation in exposure due to social distancing and other factors. Following the latent period (with duration $\delta^{-1}$), exposed people become infectious and are subsequently removed at rate $\gamma$, at which point they no longer infect others. 
A proportion $\iota$ (the infection fatality rate, or IFR) of removed individuals die from COVID at temporal rate $\mu$, and the rest remain alive. 

As a simplifying approximation, our model assumes a conserved population, i.e., there are no births and no deaths due to competing risks:
\begin{linenomath*}
\[
S(t)+E(t)+I(t)+R_S(t)+R_D(t)+D(t) = N
\]
\end{linenomath*}
for all times $t$, where $N$ is the state's total population.
Note that the time-varying basic reproduction number $R_0(t)$ and effective reproduction number $R_e(t)$, which describe rates of transmission in the initial and current population, respectively, are given by $R_0(t) = \beta(t)/\gamma$ and $R_e(t)=S(t)R_0(t)$.

We assume that $\gamma^{-1}$, the average length in days of the infectious period, is determined by the disease and constant over time. 
We make the same assumption for the other biological parameters introduced above.
In particular, while the IFR $\iota$ can realistically change over time, e.g., due to vaccination, the time period of our study focuses on viral transmission prior to widespread vaccine administration and circulation of novel SARS-CoV-2 strains 
with differential virulence. Estimates of the IFR over time in England based on regular random testing of the population found that, while the IFR did fluctuate in 2020, it hovered around 0.67\% \citep{ifr-england}. This is consistent with the IFR estimated in a systematic meta-analysis in 2020 \citep{ifr-meta}, with the results of \citet{irons}, and with our estimates discussed in Section \ref{sec:results}. 

As another simplifying approximation, our model does not account for waning immunity and reinfection, as acquired immunity was relatively long-lasting and reinfection within the first year of the pandemic was rare \citep{hall-immunity,pooley-immunity,helfand-immunity,ma2023reinfection,deng2022reinfection,medic2022reinfection,flacco2022reinfection,guedes2023reinfection,murchu2022reinfection,racine2022reinfection}. As those with prior infection were subject to a lower risk of death, this simplification circumvents the need to model the reduced IFR among reinfections.

Regarding prior specification, $R_0(t)$ is given a scaled beta-distributed random walk structure. We assume that $R_0(t)$ is constant during each week and, in an abuse of notation, write $R_0(w(t))$ to mean the value of $R_0$ in week $w(t)$ to which day $t$ belongs. We have
\begin{linenomath*}
\begin{align*}
R_0(w+1)/R_0^{\max} &\sim \text{Beta}(\sigma^2_R R_0(w)/R_0^{\max}, \sigma^2_R(1-R_0(w)/R_0^{\max})), \\
R_0(0) &\sim \text{Uniform}(0,R_0^{\max}), \\
\pi(\log\sigma^2_R) &\propto 1.
\end{align*}
\end{linenomath*}
The prior on $R_0(w+1)$ is centered at $R_0(w)$.
We place a flat improper prior on the log-transformed scale parameter 
$\log\sigma^2_R$.
We take $R_0^{\max}=6.5$ to be the upper bound for the transmission rate based on \citep{liu-r0}.
We place a flat Dirichlet prior on the initial SEIRD components:
\begin{linenomath*}
\begin{align*}
S(0) &= (1-p) + p\cdot x_0(S), \\
E(0) &= p\cdot x_0(E), \\
I(0) &= p\cdot x_0(I), \\
R_S(0) &= p\cdot(x_0(R)+x_0(D))(1-\iota), \\
R_D(0) &= p\cdot x_0(R)\cdot\iota, \\
D(0) &= p\cdot x_0(D)\cdot\iota, \\
(x_0(S),x_0(E),x_0(I),x_0(R),x_0(D)) &\sim \text{Dirichlet}(1,1,1,1,1),
\end{align*}
\end{linenomath*}
Here $p = 0.05$ is the upper bound on the proportion of the population potentially infected at or before time 0 (the first day of the study period). The remaining parameters are detailed in Table \ref{tab:epi-params}. We specify state-specific priors for the IFR using a normal distribution truncated to the unit interval based on the posterior median and 95\% credible interval reported by \citet{irons}:
\begin{linenomath*}
\[
\iota_s \sim \text{Normal}_{[0,1]}(\hat\iota_s,\hat\sigma_{s}).
\]
\end{linenomath*}
Here $\hat\iota_s$ is the posterior median IFR in state $s$ and $\hat\sigma_s$ is obtained by dividing the width of the 95\% credible interval by 4 (the ``range over 4'' rule).

\begin{table}[t!]
\centering
\caption{Epidemiological parameters. All times are in days. 
}
\begin{tabular}{|p{60mm}|p{20mm}|p{50mm}|}
\toprule
\textbf{Parameter}  & \textbf{Value} & \textbf{Reference} \\
\midrule
$\delta^{-1}$: Mean duration of latent period & 5.5 & 
\citet{xin2022latent}\newline
\citet{linton-incubation}\newline
\citet{lauer-incubation}\newline
\citet{gallo-parameters}\newline
\citet{wu-incubation}
\\
\hline 
$\gamma^{-1}$: Mean duration of infectious period & 5.0 &
\citet{hakki2022gamma}
\\
\hline
$\mu^{-1}$: Mean time to death after removal & 10.5 & 
\citet{linton-incubation}\newline
\citet{byrne-infectious} \\
\hline
$\tau_D$: Mean time from exposure to death & 21.0 & \citet{ward-death-delay}\\
\hline
$\iota_s$: State-specific IFR & 
Varying
& \citet{irons}\\
\hline
$R_0^{\max}$: Upper bound on $R_0(t)$ & 6.5 & \citet{liu-r0}\\
\hline 
Mean time from case reporting to death & 8.053 & \citet{jin-ccd} \\
\hline 
Standard deviation in time from case reporting to death & 4.116 & \citet{jin-ccd} \\
\bottomrule
\end{tabular}
\label{tab:epi-params}
\end{table}

\subsubsection{Likelihood on deaths}
\label{sec:deaths}

\begin{figure}
    \centering
    \includegraphics[width=\textwidth]{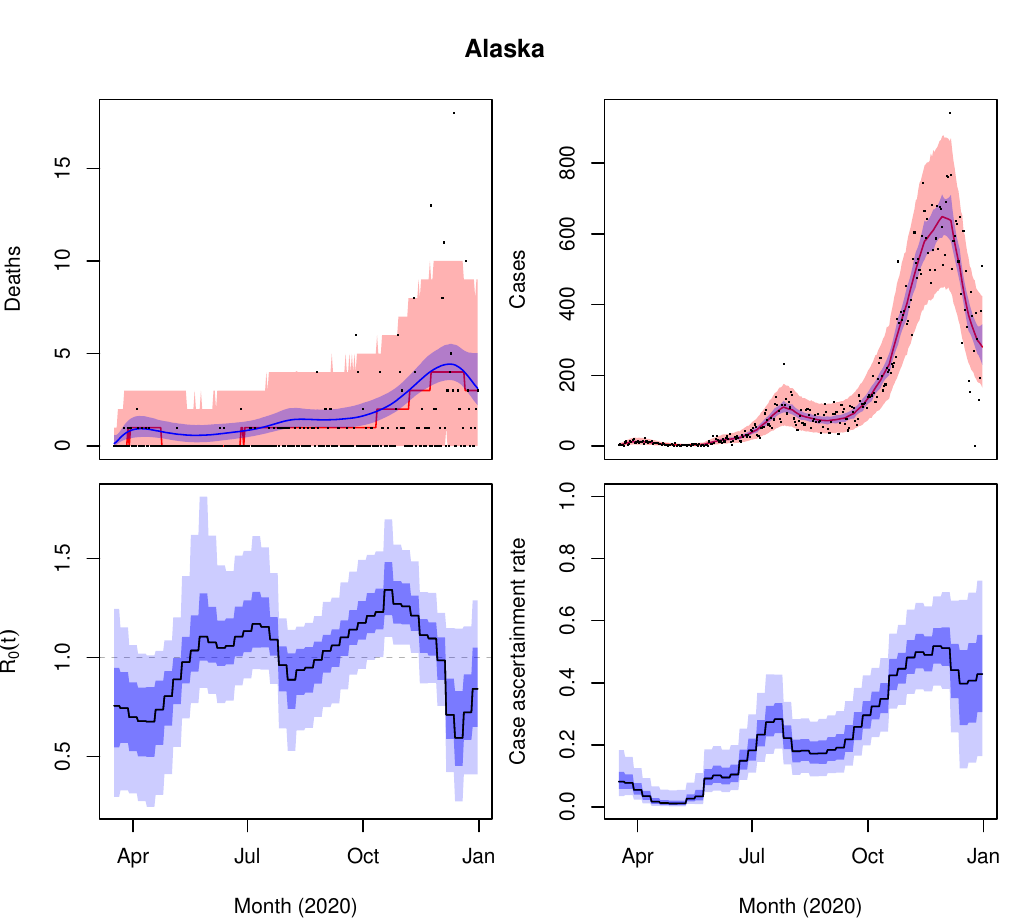}
    \caption{SEIRD model fit to COVID data in Alaska. \textbf{Top panels:} observed deaths $d(t)$ and cases $c(t)$ are plotted in black. Median and 90\% credible intervals of the posterior predictive distributions of $d(t)$ and $c(t)$ are in red. Posterior median and 90\% credible intervals of the underlying mean parameters $m_D(t)$ and $m_C(t)$ are in blue.
    \textbf{Bottom panels:} Posterior median, 50\%, and 90\% credible intervals for the basic reproduction number $R_0(t)$ and the case ascertainment rate $\text{CAR}(t)$.
    }
    \label{fig:alaska}
\end{figure}

In a given U.S. state, let $d(t)$ and $c(t)$ denote the number of COVID deaths and cases recorded in the state on day $t$, as recorded the JHU CSSE \citep{jhu-csse-data}. To account for measurement error, idiosyncratic reporting, and overdispersion in viral transmission \citep{overdispersion1,overdispersion2,overdispersion3,overdispersion4,overdispersion5,overdispersion6}, we use a zero-inflated negative binomial model on $d(t)$ and $c(t)$. 
Many states inconsistently reported cases and deaths, often taking breaks over weekends and holidays, resulting in numerous spurious zeros in the data. We address this by assuming that any deaths or cases occurring on such a day are reported on the first subsequent day of accurate reporting.  

Specifically, let $Z_t$ indicate the event that the number of deaths occurring on day $t$ is incorrectly reported as 0. We assume that the $Z_t$ are independent and identically distributed with $P(Z_t=1)=\theta_D$. We know that $Z_t=0$ on days with reported deaths ($d(t)>0$) and our model conditions on this knowledge. Assume $t_0<t_0+k$ are days with $d(t_0)>0$ and $d(t)=0$ for all $t\in(t_0,t_0+k]$. Note that some of the zeros on days $t\in(t_0,t_0+k]$ could be due to misreporting, whereas others could be accurate reporting days on which zero deaths actually occurred. We marginalize over the unknown random variables $Z_t$ for $t\in(t_0,t_0+k]$ conditional on the assumptions that: the reported deaths $d(t_0)$ are centered at $m_D(t_0) = N\mu R_D(t_0)$, the true number of deaths on day $t_0$; in expectation, any deaths occurring on misreporting days are reported on the next day of accurate reporting. Under these assumptions, the underlying mean of observed cases $d(t_0+k)$ conditional on $Z_{t_0+k}=0$ is, after marginalizing over $Z_t,t\in(t_0,t_0+k)$,
\begin{linenomath*}
\begin{align*}
m_D(t_0+k) &= \mathbb{E}[d(t_0+k)|Z_{t_0+k}=0] \\
&= \mathbb{E}\left[m_D(t_0+k)+\sum_{t\in(t_0,t_0+k)}Z_t m_D(t)\right] \\
&= \sum_{t=0}^{k-1} m_D(t_0+k-t)\theta_D^{t}.
\end{align*}
The likelihood on observed deaths is then given by the following zero-inflated negative binomial:
\begin{align}
P\left(d(t)=d\ |\ m_D(t),\kappa_D(t),\theta_D\right) &= \begin{cases}
\theta_D +(1-\theta_D)\cdot\text{NegBin2}(0,\kappa_D(t)^{-1}), & d=0, \\
(1-\theta_D)\cdot\text{NegBin2}(m_D(t), \kappa_D(t)^{-1}), & d>0,
\label{eq:pmf-deaths}
\end{cases}
\end{align}
\end{linenomath*}
where NegBin2$(\mu,\tau)$ is parametrized to have mean $\mu$ and variance $\mu+\mu^2/\tau$. We allow the overdispersion parameter $\kappa_D(t)$ to depend on the mean as follows:
\begin{linenomath*}
\[
\kappa_D(t)^{-1} = \kappa_D^{-1}\left(\zeta_D m_D(t) + (1-\zeta_D)\right),
\]
\end{linenomath*}
where $\zeta_D\in[0,1]$ is a proportion parameter and $\kappa_D\in(0,\infty)$. We use $\text{Uniform}(0,1)$ priors on $\theta_D$ and $\zeta_D$ and a flat improper prior $\pi(\log\kappa_D)\propto 1$. Finally, with $\tilde{d}(0)$ representing the cumulative deaths reported prior to the start of the modeling window, we use the likelihood
\begin{linenomath*}
\[
\tilde{d}(0)\sim\text{Poisson}(N\cdot (R_D(0)+D(0))).
\]
\end{linenomath*}

This model flexibly interpolates between a count distribution with a linear mean-variance relationship (as with the overdispersed Poisson) when $\zeta_D=0$ and a quadratic mean-variance relationship (as with the usual negative binomial) when $\zeta_D=1$. We found that this modification was necessary to accurately capture dispersion in clinical data across a range of states. In some states, such as Ohio and Indiana, a standard Poisson was sufficient to produce well-calibrated posterior predictive distributions. In most other states, such as Texas and Florida, a negative binomial was required. Finally, there were some states, such as New York, in which Poisson predictive intervals were too narrow and negative binomial intervals were too wide, while predictive intervals derived from the model \eqref{eq:pmf-deaths} were much better calibrated. Our model \eqref{eq:pmf-deaths} can handle all of these cases. Figure \ref{fig:alaska} demonstrates the model's fit to COVID data in Alaska, which exhibit large and time-varying patterns of overdispersion and zero-inflation. The mean $m_D(t)$ of the likelihood on deaths provides a much smoother representation of the data and depicts more consistent trends in transmission reflected in the case data. 


\subsubsection{Likelihood on cases}

Let $\nu(t)=N\beta(t)S(t)I(t)$ denote the number of new infections in the state on day $t$. We relate the true prevalence $\nu(t)$ to the number of cases $c(t)$ reported on each day using a compartmental model that accounts for time-varying imperfect case ascertainment and delays between exposure and case confirmation via testing. We define a ``number of infections waiting to be confirmed'' compartment $I_C(t)$ satisfying
\begin{linenomath*}
\[
I_C(t+1) = I_C(t)(1-\tau) + \text{CAR}(t+1)\nu(t+1),
\]
\end{linenomath*}
where $\tau^{-1}$ is the expected delay in days from infection to case confirmation and $\text{CAR}(t)$ is the case ascertainment rate on day $t$. We use a truncated normal prior on the case confirmation delay based on \citep{jin-ccd}: 
\begin{linenomath*}
\[
\tau \sim N_{[\tau_D-8.053-1.96\cdot 4.116,\tau_D]}(\tau_D - 8.053,4.116),
\]
\end{linenomath*}
where $\tau_D$ is the mean total time from infection to death 
\begin{linenomath*}
\[
\tau_D=\delta^{-1}+\gamma^{-1}+\mu^{-1} = 21.0.
\]
\end{linenomath*}
The underlying mean $m_C(t)$ of $c(t)$ on misreporting days $t=t_0+k$ is analogous to that for deaths, $m_D(t)$, with the expected number of deaths on accurately reported days $t_0$, $N\mu R_D(t_0)$, replaced by the expected number of infections confirmed on day $t_0$, $\tau I_C(t_0)$:
\begin{linenomath*}
\begin{align*}
m_C(t_0) &= \tau I_C(t_0), \\
m_C(t_0+k) &= \sum_{t=0}^{k-1} m_C(t_0+k-t)\theta_C^{t}.
\end{align*}
\end{linenomath*}
The zero-inflated negative binomial likelihood is then
\begin{linenomath*}
\begin{align*}
P\left(c(t)=c\ |\ m_C(t),\kappa_C(t),\theta_C\right) &= \begin{cases}
\theta_C + (1-\theta_C)\cdot\text{NegBin2}(0;m_C(t),\kappa_C(t)^{-1}), & c=0, \\
(1-\theta_C)\cdot\text{NegBin2}(c;m_C(t),\kappa_C(t)^{-1}), & c>0, \\
\end{cases} \\
\kappa_C(t)^{-1} &= \kappa_C^{-1}\left(\zeta_C m_C(t) + (1-\zeta_C)\right).
\end{align*}
\end{linenomath*}
We use $\text{Uniform}(0,1)$ priors on $\theta_C$ and $\zeta_C$ and flat improper priors $\pi(\log\kappa_C)\propto 1$ and $\pi(\log I_C(0))\propto 1$.
We place a beta-distributed random walk prior on case ascertainment rates:
\begin{linenomath*}
\begin{align*}
\text{CAR}(0) &\sim \text{Uniform}(0,1), \\
\text{CAR}(t+1) &\sim \text{Beta}(\sigma^2_{\text{CAR}}\text{CAR}(t),\sigma^2_{\text{CAR}}(1-\text{CAR}(t))), \\
\pi(\log \sigma^2_{\text{CAR}}) &\propto 1.
\end{align*}
\end{linenomath*}

\subsection{Effects of NPIs on SARS-CoV-2 transmission}
\label{sec:bhm}

We now turn to the regression model linking NPI policies to the dynamics of viral transmission. Our main source of data is OxCGRT, which aggregates and continuously updates national and subnational government policy responses to the pandemic at the daily level starting from January 1, 2020 \citep{oxcgrt}. The database tracks a range of containment and closure, economic, health, and vaccination indicators with numerical values corresponding to the strength of the response on each day. Our transmission regression model focuses on the following 11 policy indicators: school closure, workplace closure, public event cancellation, restrictions on gatherings, public transport closure, stay-at-home requirements, restrictions on internal movement, public information campaigns, testing, contact tracing, and facial covering policies.\footnote{Notably, our model does not include international travel restrictions because: they were a policy held constant in place over time for most of 2020 (hindering identification of their effect on transmission); they were a federal policy (not relevant for state-level decision-making); and they were shown not to be very effective in reducing transmission, only delaying introduction of the virus for a few days \citep{chinazzi-travel-effect}.} To account for potential seasonality of SARS-CoV-2 transmission \citep{seasonal1,seasonal2,seasonal3}, we also included average daily temperature measurements reported for the largest population centers in each state, accessed using Meteostat Python \citep{meteostat}, as a covariate. However, models with temperature as a covariate were excluded based on model selection carried out via leave-one-out cross validation (LOO-CV) \citep{loo-cv}.

For each US state $s$, we define $u^{(s)}(t)$, an 11-dimensional vector with entries in the interval $[0,1]$ denoting the strength of each NPI implemented on day $t$. So $u_k^{(s)}(t) = 0$ represents no restrictions associated to the $k$th NPI on day $t$ (e.g., no school closure), whereas $u_k^{(s)}(t) = 1$ represents the strictest restrictions (e.g., full school closure). NPI implementation during the pandemic was highly correlated, which poses a challenge to teasing apart the effects of individual NPIs on SARS-CoV-2 transmission. We utilize a Bayesian hierarchical model (BHM) to jointly model the time-varying basic reproduction number $R_0^{(s)}(w)$ in each US state $s$ as a function of NPIs using the output from the first stage of estimation described in Section \ref{sec:epi}. The BHM leverages spatiotemporal variation in NPI implementation over time across states in order to estimate their effects. It allows for spatial heterogeneity in NPI effects (e.g., due to differential adherence to government mandates) while enabling identification via partial pooling of information across the country.

\subsubsection{Propagating uncertainty}

Let $\mathcal{D}$ denote the observed case and death data, $\mathcal{U}$ the NPI data, $\mathcal{S}$ the SEIRD model parameters, and $\theta$ the NPI regression model parameters. Our two-stage estimation procedure appropriately propagates uncertainty such that the resulting estimates approximate the posterior distribution $\pi(\theta|\mathcal{D},\mathcal{U})$ that would be obtained from combining the epidemiological and regression stages into a single model. Indeed, we have
\begin{linenomath*}
\begin{align*}
\pi(\theta|\mathcal{D},\mathcal{U}) 
&= \int \pi(\theta,\mathcal{S}|\mathcal{D},\mathcal{U}) d\mathcal{S} \\
&= \int \pi(\theta|\mathcal{S},\mathcal{D},\mathcal{U})\pi(\mathcal{S}|\mathcal{D},\mathcal{U}) d\mathcal{S} \\
&= \int \pi(\theta|\mathcal{S},\mathcal{D},\mathcal{U})\pi(\mathcal{S}|\mathcal{D}) d\mathcal{S} \tag{SEIRD parameters can be inferred from clinical data $\mathcal{D}$ alone} \\
&\approx \frac{1}{M}\sum_{i=1}^M \pi(\theta|\mathcal{S}^{(i)},\mathcal{D},\mathcal{U}),
\end{align*}
\end{linenomath*}
where 
$\mathcal{S}^{(i)}$
denotes the $i$th of $M$ posterior trajectories of the SEIRD parameters (e.g., the time-varying basic reproduction number $R_0^{(s)}(w)$ in each state $s$) derived from the first-stage transmission model posterior $\pi(\mathcal{S}|\mathcal{D})$. Here we use $M=100$ randomly sampled trajectories of $R_0^{(s)}(w)$, which define the dependent variable in the NPI regression model. For each $i=1,\ldots,M$ we generate samples from the NPI model posterior $\pi(\theta|\mathcal{S}^{(i)},\mathcal{D},\mathcal{U})$. We then aggregate these samples to obtain our final estimate of the full posterior $\pi(\theta|\mathcal{D},\mathcal{U})$.

\subsubsection{NPI regression model}
\label{sec:model:regression}

\begin{figure}[t!]
    \centering
    \includegraphics[width=\textwidth]{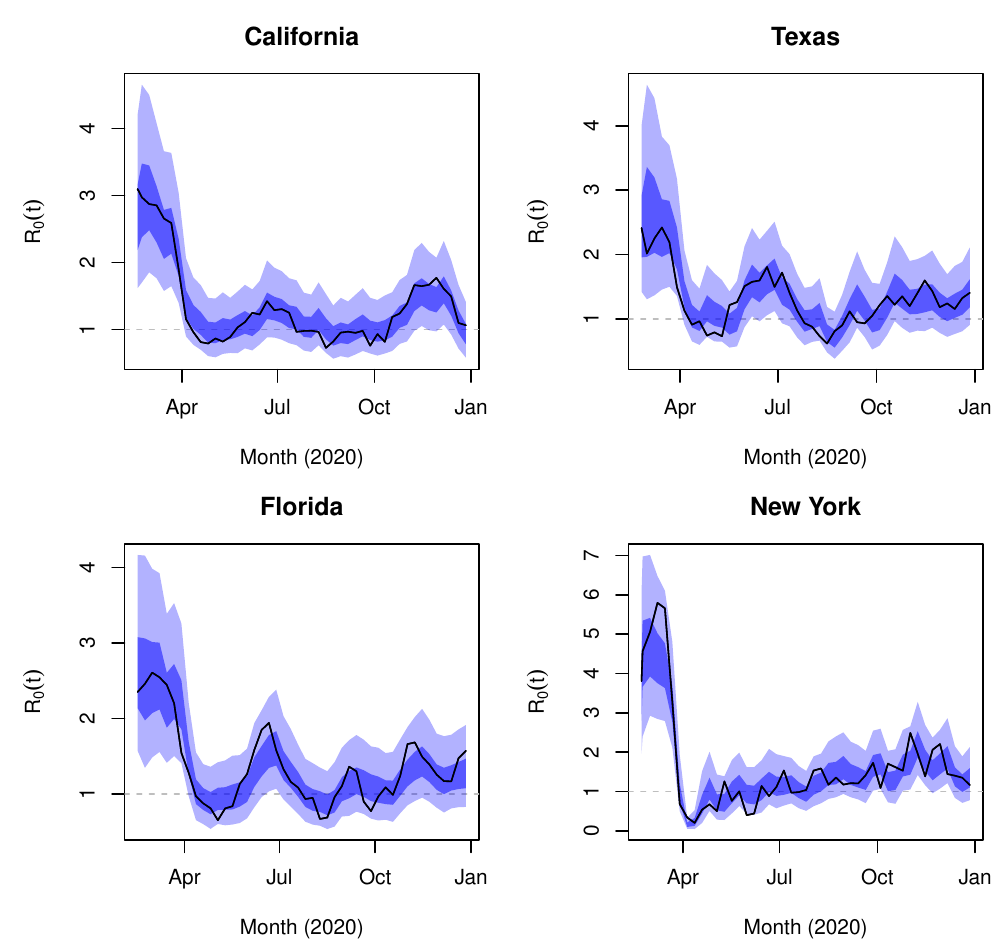}
    \caption{Time-varying transmission rates in four states. In black, the MAP trajectory output by the epidemiological model. In dark and light blue, respectively, the 50\% and 90\% credible intervals of the posterior predictive distribution from the NPI model fitted to this trajectory.}
    \label{fig:rt}
\end{figure}

Our regression model targeting the posterior $\pi(\theta|\mathcal{S},\mathcal{D},\mathcal{U})$ expresses the time-varying basic reproduction number $R_0^{(s)}(w)$ in each state $s$ and week $w$ as a log-linear function of the NPIs implemented in that week, $u^{(s)}(w)$, where we obtain weekly values for the NPIs by averaging over days.
To account for the endogenous behavioral response to the fear of infection, we also control for the expected population proportion of deaths $\tilde{D}^{(s)}(w-1)$, removals $\tilde{R}^{(s)}(w-1)$, and infections $\tilde{I}^{(s)}(w-1)$ incident in the prior week:
\begin{linenomath*}
\begin{align*}
\tilde{I}^{(s)}(w-1) &= \sum_{t:w(t)=w-1} \delta E^{(s)}(t), \\   
\tilde{R}^{(s)}(w-1) &= \sum_{t:w(t)=w-1} \gamma I^{(s)}(t), \\ 
\tilde{D}^{(s)}(w-1) &= \sum_{t:w(t)=w-1} \mu R_D^{(s)}(t).
\end{align*}
\end{linenomath*}
We consider three models: (i) only controlling for deaths $\tilde{D}^{(s)}(w-1)$; (ii) controlling for removals $\tilde{R}^{(s)}(w-1)$ and deaths $\tilde{D}^{(s)}(w-1)$; and (iii) controlling for infections $\tilde{I}^{(s)}(w-1)$, removals $\tilde{R}^{(s)}(w-1)$, and deaths $\tilde{D}^{(s)}(w-1)$.
Given the delay in case confirmation, model (ii), in which individual behavior responds to deaths and removals but not infections incident in the prior week, may be preferred on theoretical grounds. Nevertheless, our qualitative findings are consistent across models, with the main distinction being that controlling for more effects tends to attenuate the effect of school closure on transmission rates.

For notational convenience, we first define the linear predictor 
\begin{linenomath*}
\begin{align*}
\log \hat{R}_0^{(s)}(w) &:= \log R_0^{(s)} + \beta^{(s)}_u\cdot u^{(s)}(w) \\
&\qquad+ \beta^{(s)}_I\tilde{I}^{(s)}(w-1) +
\beta^{(s)}_R\tilde{R}^{(s)}(w-1) +
\beta^{(s)}_D\tilde{D}^{(s)}(w-1),
\end{align*}
\end{linenomath*}
where $R_0^{(s)}$ is the initial state-specific basic reproduction number under no restrictions, $\beta_u^{(s)}$ is a vector of state-specific random NPI effects of size $p=11$, 
$\beta^{(s)}_D$ denotes the random effect of deaths,
and similarly for $\beta^{(s)}_I,\beta^{(s)}_R$. 
Models (i) and (ii) assume $\beta^{(s)}_I=\beta^{(s)}_R=0$ and $\beta^{(s)}_I=0$, respectively.
Our final model on the observed transmission rate $R_0^{(s)}(w)$, which is output by the SEIRD model, also includes an autoregressive component:
\begin{linenomath*}
\begin{equation}
\log R_0^{(s)}(w) = \log \hat{R}_0^{(s)}(w) + \varphi\left(\log R_0^{(s)}(w-1)-\log\hat{R}_0^{(s)}(w-1)\right) + \varepsilon^{(s)}(w), 
\label{eq:rt-npi}
\end{equation}
\end{linenomath*}
where $\varphi$ is the AR(1) parameter\footnote{We considered AR($q$) models for $q=1,2,3$. We selected $q=1$ via LOO-CV.} and $\varepsilon^{(s)}(w)$ is a Student-$t$ distributed error term with $\nu_\varepsilon$ degrees of freedom:
\begin{linenomath*}
\[
\varepsilon^{(s)}(w)\sim \text{Student-}t(\nu_\varepsilon,\sigma^2_\varepsilon).
\]
\end{linenomath*}
The error terms account for unpredictable and heavy-tailed exogenous shocks that may have sustained effects on transmission (as modeled through the AR(1) term), such as the start of a new wave due to a super-spreader event or the introduction of 
new infections from an external source (e.g., due to travel into the state). The use of Student-$t$ distributed errors ensures that the regression model is robust to outliers, which prevents overfitting the effects of NPIs to the transmission data. We use a flat improper prior on $\log\nu_\varepsilon$ and find that the posterior of $\nu_\varepsilon$ concentrates between 2 and 3, indicating that a heavy-tailed error distribution is appropriate. 

We control for deaths $\tilde{D}$, removals $\tilde{R}$, and infections $\tilde{I}$ incident in the prior week following the identification strategy of a number of other studies estimating the causal effects of NPIs \citep{goolsbee-fear,deb-effects,coibion-macro,crucini-orders,chernozhukov-masks,verschuur-shipping,mader-npis,KARAIVANOV-mask-effect}. The existence of substantial voluntary social distancing and its pronounced economic effects in the US and elsewhere have been well-documented in numerous empirical analyses \citep{bartik-labor,bodenstein-social,goolsbee-fear,aum-korea,chen-impact,chernozhukov-masks,jamison-behavior,badr-mobility,abouk-behavior,cronin-social} and derived from first principles in macroeconomic modeling \citep{eichenbaum-macro,farboodi2021,brzezinski-voluntary,krueger-swedish}. In response to SARS-CoV-2 outbreaks, people began social distancing (and, potentially, other protective measures, e.g., mask-wearing) prior to the onset of restrictions and subsequently increased social activity separate from the lifting of restrictions. As a result, declines in mobility, consumer spending, and hours worked cannot be fully attributed to the effects of NPI policies \citep{forsythe-labor,baker-spending}. Including prior deaths, removals, and infections as covariates in the regression model accounts for changes in 
protective 
behaviors by individuals responding to the risk of infection.\footnote{This model was selected by LOO-CV among models controlling for both deaths and cases in the past $x$ weeks, where $x$ was fixed at $1,2,4,6,8,10,12,16$ and also allowed to vary as a parameter in the model.} 

Figure \ref{fig:dag} depicts a directed acyclic graph (DAG) representing our causal model for the outcome $R_0(w)$ in each week. We suppress the state $s$ for compactness of notation. Deaths, removals, and infections incident in week $w-1$ may affect the policy response $u(w)$ and the transmission rate $R_0(w)$ in the following week, with the latter effect representing the endogenous behavioral response (including social distancing and other protective measures) to the fear of infection. The transmission rate $R_0(w)$ is also a function of NPI policies $u(w)$ and exogenous shocks $\varepsilon(1:w):=\{\varepsilon(v):v=1,\ldots,w\}$.
We allow for past shocks $\varepsilon(1:w-1)$ to affect past deaths, removals, and infections and the current policy response $u(w)$. In our regression model \eqref{eq:rt-npi}, the lagged and attenuating effect of past shocks on the transmission rate is captured by the AR(1) term:
\begin{linenomath*}
\[
\varphi\left(\log R_0^{(s)}(w-1)-\log\hat{R}_0^{(s)}(w-1)\right) = \sum_{v=1}^{w-1} \varphi^{w-v}\varepsilon^{(s)}(v).
\]
\end{linenomath*}
Given the DAG \ref{fig:dag}, we see that controlling for $(\tilde{I}(w-1),\tilde{R}(w-1),\tilde{D}(w-1))$ and $\varepsilon(1:w-1)$---as we do in \eqref{eq:rt-npi}---blocks all back-door paths from $u(w)$ to $R_0(w)$. As such, the effect of NPIs is identified in this model following the back-door criterion \citep{pearl-causality}.

Regarding prior specification for the regression \eqref{eq:rt-npi}, we use a hierarchical model for the state-specific coefficients $\theta^{(s)}=(\beta^{(s)}_u,\beta^{(s)}_I,\beta^{(s)}_R,\beta^{(s)}_D,R_0^{(s)})\in\mathbb{R}^{p+4}$, which enables partial pooling of information. With $\theta$ denoting the global pooled effects, we have
\begin{linenomath*}
\begin{align}
\theta^{(s)} &\sim \text{Normal}(\theta,V)\prod_{k=1}^p I\left(\theta^{(s)}(k)\le 0\right), 
\label{eq:npi-prior}
\\
V &= D(\lambda)\Omega D(\lambda), \nonumber\\
\Omega &\sim \text{LKJ}(\zeta=1), \nonumber\\
\lambda(j) &\sim \text{Student-}t_{[0,\infty)}(0,2.5^2,3), \quad j=1,\ldots,p+4, \nonumber\\
\sigma_\varepsilon &\sim \text{Student-}t_{[0,\infty)}(0,2.5^2,3), \nonumber \\
\pi(\log\nu_\varepsilon) &\propto 1,\nonumber \\
\varphi &\sim \text{Uniform}(-1,1). \nonumber
\end{align}
\end{linenomath*}
The truncated normal prior \eqref{eq:npi-prior} on the state-level random effects $\theta^{(s)}$ assumes that they are centered around the pooled effects $\theta$ and that NPIs cannot increase the transmission rate ($\theta^{(s)}(k)\le 0$), in line with the results of numerous studies estimating the effects of NPIs \citep{haug-npi-effects,sharma-effects,brauner-effects,flaxman-effects,hsiang-npi-effects,liu-npi-effects,li-npi-effects,jamison-behavior,stokes-npi-effects,banholzer-npi-effect}. 
For the remaining parameters, we use the default prior specification for multilevel models used in the brms R package \citep{brms}. The covariance matrix $V$ of the random effects is decomposed as the product of a correlation matrix $\Omega$ given an LKJ prior with parameter $\zeta=1$ (specifying a uniform prior on correlation matrices), and a diagonal matrix $D(\lambda)$ with entries $\lambda(j)$ given Student-$t$ priors with 3 degrees of freedom truncated to be non-negative. The error standard deviation $\sigma_\varepsilon$ is given the same truncated Student-$t$ prior. The log degrees of freedom $(\log\nu_\varepsilon)$ for the Student-$t$ error terms $\varepsilon^{(s)}(w)$ is given a flat improper prior. The AR(1) parameter $\varphi$ is given a $\text{Uniform}(-1,1)$ prior. 

\subsection{Evaluating and optimizing costs}
\label{sec:methods-cost}

The SEIRD equations \eqref{eq:seird} combined with the NPI regression model \eqref{eq:rt-npi} define a simulator for the trajectories of infections and deaths under counterfactual NPI policies, conditional on the parameters estimated using the NPI and clinical data---denoted $\mathcal{U}$ and $\mathcal{D}$, respectively. Evaluating the cost-effectiveness of NPIs and determining optimal strategies requires accounting for the aggregate costs incurred in implementing policies and their consequent health impacts. For an NPI policy $\mathbf{u}=\{u(t)\}_{t=1}^T$ implemented in a state on the days $t=1,\ldots,T$, we define its associated cost $\mathcal{C}(\mathbf{u})$ as the sum of the posterior expected costs incurred by infections and NPI implementation:
\begin{linenomath*}
\begin{equation}
\mathcal{C}(\mathbf{u}) = \mathbb{E}\left[c_{\text{NPI}}(\mathbf{u})+
\frac{1}{N}
\sum_{t=1}^T c_\nu \nu(t)\bigg|\mathcal{D},\mathcal{U}\right],
\label{eq:cost}
\end{equation}
\end{linenomath*}
where $\nu(t)=N\beta(t)S(t)I(t)$ is the number of new COVID infections in the state incident on day $t$ under the policy $\mathbf{u}$. Here $c_\nu$ is the average cost in USD2020 associated to a COVID infection and 
\begin{linenomath*}
\[
c_{\text{NPI}}(\mathbf{u}) = \sum_{k=1}^p c_k(\mathbf{u}_k)
\]
\end{linenomath*}
is the average per capita cost in USD2020 associated to implementing the policy $\mathbf{u}=(\mathbf{u}_1,\ldots,\mathbf{u}_p)$, where $c_k$ is the cost of the $k$th NPI.  We define these quantities below.\footnote{Given the relatively short duration of our study period (i.e., the first year of the pandemic), we do not adjust the cost function for temporal discounting.} Note that, in practice, we cannot directly evaluate the expectation \eqref{eq:cost}. Instead, we use posterior trajectories of infections $\nu(t)$ to approximate \eqref{eq:cost} via Monte Carlo.

\subsubsection{Cost of infections}
\label{sec:cost-infection}

The average cost of a COVID infection is a sum of average life costs (due to COVID deaths), medical costs (incurred by treatment), productivity costs (due to worker absenteeism), and costs associated to voluntary social distancing in response to the fear of infection.

The life cost associated to a COVID infection deserves some discussion, as it dominates the aggregate cost of infection and it requires ascribing a dollar value to death, which can be a contentious issue. In cost-benefit analysis, the standard approach to quantifying mortality risk reductions as a result of public policy in monetary terms is through the value of a statistical life (VSL), commonly estimated at about \$11 million in USD2020 \citep{greenstone-distancing}. In our context, the relevant quantity is the value of a statistical COVID death (VSCD). As \citet{robinson-vsl} note, COVID deaths are concentrated in the oldest age groups and the decision to adjust the VSCD for the age profile of COVID mortality or not can vastly alter the conclusions of a cost-benefit analysis. As such, we conduct sensitivity analysis of our results using high and low estimates of the VSCD reported in \citep{robinson-vsl}. 
In line with a number of other studies in the COVID economics literature \citep{farboodi2021,hall-vsl,greenstone-distancing,kaplan-covid-political-economy}, we use as our baseline the low estimate of \$4.47 million, which is based on the average years of life lost to a COVID death and a constant value per statistical life year.\footnote{That is, outside of our sensitivity analysis (Section \ref{sec:sensitivity}), all estimated costs are based on the low VSCD.} 
The high estimate of \$10.63 million assumes the VSCD equals the population average VSL (i.e., it does not adjust the VSL for the age pattern of COVID deaths). Table \ref{tab:econ-params} records the value of this and other economic parameters used in our study. Denoting the VSCD by $c_{\text{VSCD}}$, the average life cost per COVID infection in state $s$ is then $c_{\text{VSCD}}\cdot\tilde\iota^{(s)}$, where $\tilde\iota^{(s)}$ is the posterior average state-level IFR. 

To quantify the cost associated to voluntary social distancing, we rely on the results of \citet{aum-korea}, who find that a one per thousand increase in the COVID case rate caused a 2.68\% drop in employment in South Korea in the spring of 2020, with similar (although non-causal) estimates in the US and UK. We adjust their numbers for under-reporting of cases based on the number of deaths and cases in South Korea in their period of study reported by Our World in Data \citep{owidcoronavirus}. By February 29, 2020, South Korea had 556 cumulative confirmed cases, and by March 7 it had 3,526. The cumulative COVID deaths three weeks later on March 21 were 75, and by March 28 were 104. With an IFR of 0.68\% \citep{ifr-meta}, we would expect between 11,029 and 15,294 infections with this number of deaths. This suggests a case ascertainment rate between 5\% and 23\% in that period. We average these two numbers, assuming 15.5\% case ascertainment, which yields a $0.155\cdot 2.68 \approx 0.42\%$ drop in employment resulting from a one per thousand increase in COVID prevalence. In line with \citet{chetty-impacts}, \citet{aum-korea} find that these impacts were felt most acutely among low-wage workers. As such, we phrase this increase in unemployment due to an infection---which is assumed to last the average duration of the infectious period ($\gamma^{-1}=5.0$ days)---in monetary terms using the state's median income.\footnote{We use national and state-level personal income data reported by \citet{bea-expenditure}. To approximate state-level median income, we multiply the US median personal income by the state's per capita income divided by the US per capita income.} In sum, we find that the fear cost per COVID infection ranges from \$1,491 to \$3,199 across states with a median of \$1,999.

For the remaining parameters, we take \$3,045 as the average medical cost of a COVID infection based on the estimates of \citet{bartsch-healthcare-cost,demartino-healthcare-cost}. We assume the average productivity cost of a COVID infection is equal to one week of sick days at a state's median wage, which yields costs similar to those reported by \citet{skarp-npi-cost-review}. These range from \$505 to \$1,084 across states with a median of \$677.

\subsubsection{Cost of workplace closures and social distancing measures}
\label{sec:cost-social}

Beginning in the spring of 2020, consumer spending dropped significantly in response to health concerns and government-mandated business closures and social distancing policies. This reduction in consumer spending---primarily on in-person services---was responsible for a large majority of the decline in US GDP in the second quarter of 2020. Declining business revenue led to substantial layoffs with subsequent unemployment increases concentrated among low-wage workers \citep{chetty-impacts}. As such, we quantify the cost of workplace closures and social distancing measures through their effects on employment, which have been thoroughly studied in the COVID economics literature \citep{coibion-macro,crucini-orders,bartik-labor,bodenstein-social,alexander-orders,gupta-labor-effects,baek-unemployment,barrot-business-effect}. As above, we convert employment rate decreases in each state to monetary losses using the state's median personal income, which reflects the wage distribution of pandemic job loss. 

The effects of COVID workplace closures on unemployment and consumer spending in the US were estimated by \citet{gupta-labor-effects,barrot-business-effect,crucini-orders}, who arrive at broadly similar conclusions. 
\citet{crucini-orders} found that non-essential business closures led to a 1-2 percentage point decline in expenditures. 
\citet{gupta-labor-effects} found that 60\% of the 12 percentage point decline in the employment rate between January and April 2020 was due to state policies, with government-mandated business closures and stay-at-home orders each accounting for half of those 7.2 percentage points.\footnote{Although drops in consumer foot traffic are not directly comparable to employment rate decreases, \citet{goolsbee-fear} found that general shelter-in-place orders reduced overall consumer visits by 7 percentage points. 
} 
\citet{barrot-business-effect} found that a 10 percentage point increase in the share of restricted labor was associated with a 3 percentage point decline in April 2020 employment.
Given these findings, we define low, middle, and high scenarios in which workplace closures cause a 2\%, 4\%, and 6\% declines in the employment rate. We take the middle scenario as our baseline and consider the low and high scenarios in our sensitivity analysis in Section \ref{sec:sensitivity}.

We define social distancing measures as the combination of the following six NPIs tracked by OxCGRT: stay-at-home orders, restrictions on gatherings, restrictions on internal movement, public information campaigns, public transit closures, and public event cancellations. We bundle these policies for a number of reasons: their implementation was highly correlated in time and space; there is a paucity of information on the individual economic effects of most of these interventions as the COVID economics literature tends to focus on ``social distancing'' or ``lockdown'' measures broadly defined (likely due to their synchronous adoption); and they are blanket policies acting as relatively blunt instruments with their primary direct effects on the economy stemming from a common mechanism---namely, reduction in consumer spending on in-person services with consequent unemployment. 

The effects of social distancing measures on unemployment and consumer spending in the US were studied by \citet{coibion-macro,crucini-orders,bodenstein-social,gupta-labor-effects,baek-unemployment}.
Drawing on survey responses, \citet{coibion-macro} found that individuals in counties under lockdown were 2.8 percentage points less likely to be employed relative to other survey participants, had a 1.9 percentage point lower labor-force participation, and had a 2.4 percentage point higher unemployment rate. 
\citet{crucini-orders} found that stay-at-home orders caused a 4 percentage point decrease in consumer spending and hours worked.
\citet{bodenstein-social} found that the combined effect of voluntary and mandatory social distancing could explain 6–8 percentage points of the 12\% drop in US GDP in the second quarter of 2020 and that stay-at-home orders could account for a 2 percentage point increase in the unemployment rate.
As mentioned above, \citet{gupta-labor-effects} found that stay-at-home orders led to a 3.6 percentage point decline in employment rates through April 2020. 
Similarly, \citet{baek-unemployment} found that each week of stay-at-home order exposure between March 14 and April 4, 2020 yielded an increase in a state's weekly  unemployment insurance claims corresponding to 1.9\% of its employment level. As \citet{bartik-labor} note, nearly all employment declines occurred within the two-week period March 14--28, which implies a cumulative 3.8\% drop in the employment rate based on the findings of \citet{baek-unemployment}. Hence, we assume that social distancing measures cause a 4\% decline in the employment rate. In our sensitivity analysis, we do not vary the cost of social distancing measures as we are primarily interested in assessing the robustness of the optimal strategy and the relative costs of various policies rather than variation in the total cost incurred by each policy, which means that we are free to leave the value of one term in the cost function \eqref{eq:cost} fixed.

We note that the combined economic effects of workplace closures and social distancing measures used here are on par with trends in aggregate economic output in the US and elsewhere observed in 2020 and in prior pandemics. \citet{cbo} estimates a 3.5\% year-over-year decline in real US GDP from 2019 to 2020. Analyzing trends in annual global GDP, \citet{kaplan-covid-political-economy} estimate a 7\% decline from 2019 to 2020 due to COVID, which equates to a loss of \$10 trillion. \citet{demirguc-kunt-npi-impact} find that national lockdowns led to a 10\% decline in economic activity across Europe and Central Asia in the spring of 2020.
Studying the 1918 Spanish flu pandemic, in which social distancing measures were the primary tools used to curtail viral spread, \citet{barro-spanish-flu-econ} estimate a cumulative loss in GDP per capita of 6\% over 3 years.


\subsubsection{Cost of school closures}
\label{sec:cost-school}

The cost of school closures is a sum of productivity loss due to worker absenteeism (as parents of children out of school miss work to care for their kids) and learning loss resulting from students missing school and receiving lower quality education through distance learning. 

\citet{lempel-school-cost,sadique-school-cost} estimated the magnitude of direct GDP loss due to worker absenteeism resulting from extended school closures in the US and UK, respectively, and arrived at nearly identical numbers. They find that four weeks of school closure would cost 0.1--0.3\% of GDP in the US and 0.1--0.4\% in the UK. For our study, we use 0.2\%. Similar estimates based on modeling studies are reviewed in \citet{viner-school-review}. 

Notably, \citet{lempel-school-cost} also estimate the healthcare impacts of a four-week school closure in the US, finding that it would lead to a reduction of 6\% to 19\% in key healthcare personnel. Similarly, \citet{bayham-school-workforce} find that 15\% of the healthcare workforce would be in need of childcare during a school closure and find that their absence from work could cause a greater number of COVID deaths than school closures prevent. Pricing these health impacts is not straightforward, so we omit these considerations when defining the cost function. As such, we believe that our accounting of the costs of school-closure-related worker absenteeism is conservative.

While learning loss due to school closure can be viewed as a social cost, it can lead to substantial downstream economic costs as cohorts of students that missed significant schooling eventually enter the labor-force as less skilled and productive workers. Education economists have extensively studied the connections between time spent in school, performance on standardized tests, and subsequent impacts on lifetime earning and GDP with findings that are consistent across contexts. \citet{hanushek-learning,psacharopoulos-learning}---whose assessments of the cost of learning loss we use---provide discussion and references. As our high scenario, we use the estimate of \citet{hanushek-learning}, who find that cohort learning loss equivalent to one-third of a school-year has a staggering net present value equal to 69\% of current-year GDP.\footnote{\citet{hanushek-learning} also estimate that a student missing 0.33 years of school leads to a loss in lifetime individual income of 3.0\% in the US and 2.6\% pooled globally. \citet{fuchs-welfare-school} find average losses of 2.1\% in lifetime earnings and 1.2\% in permanent consumption of children affected by COVID-19 school closure. Considering that \citet{betthauser-learning-review} report a learning deficit of 0.35 school-years accrued during COVID, the estimates of \citet{hanushek-learning} and \citet{fuchs-welfare-school} are quite similar.}
\citet{psacharopoulos-learning} arrive at a much smaller number, finding a 9\% GDP loss arising from 0.33 years of lost schooling, which forms our low scenario and also our baseline value. We note that the results of \citet{psacharopoulos-learning} are predicated on the assumption that remote learning is 90\% as effective as in-person school, which is a likely source of the large discrepancy between the two estimates. While distance learning certainly mitigated some learning loss \citep{betthauser-learning-review}, and keeping schools open during the pandemic would have also incurred some learning loss due to student and teacher illness-related absenteeism, we believe that this assumption leads to a conservative estimate of the cost of learning loss associated to in-person school closure.\footnote{
Indeed, based on the results of a recent preprint \citep{fahle-us-learning-loss}, \citet{nyt-schools} demonstrate that drops in math scores in mostly in-person school districts were only 2/3 of those in mostly remote or hybrid districts among third through eighth graders in the U.S. during COVID-19.
} 
Nevertheless, as we discuss in Section \ref{sec:results}, we find that optimal NPI strategies based on this low estimate involve no closure of schools beyond the usual 16 weeks of break per year.

In their systematic review and meta-analysis, \citet{betthauser-learning-review} find a substantial and consistent learning deficit of 0.35 school-years of learning loss across 15 high- and middle-income countries, which accrued early in the pandemic.\footnote{
Citing \citep{fahle-us-learning-loss}, \citet{nyt-schools} note that aggregate learning loss in the U.S. during COVID-19 likely exceeded 0.35 school-years. Again, we believe that our accounting of the costs of school closure is conservative as such.
} 
This learning gap persisted but ceased to grow beyond 0.35 school-years, which suggests that remote learning did mitigate learning loss with greater efficacy (relative to in-person schooling) as time went on.\footnote{Similarly, based on results of a simulation model published earlier in the pandemic, \citet{azevedo-school-impacts} projected that COVID-19 school closures could result in learning loss equivalent to 0.3--1.1 years of schooling.} 
In our cost function, we account for the improving quality of remote learning over time by assuming that the amount of learning loss incurred by one week of school closure equates to one school-week initially and decreases linearly to 0 as a function of the cumulative number of past weeks spent under school closure, such that 0.35 school-years is the maximal cumulative amount of learning loss possible. Furthermore, our cost function only accounts for marginal learning loss (i.e., beyond what would be expected after summer break, for example) by assuming that learning loss only begins to accrue once the duration of school closure exceeds 16 weeks.

\subsubsection{Cost of testing, tracing, and masking}

We quantify the per capita cost of a week-long mask mandate as the price of supplying an individual with masks for a week. Following \citet{bartsch-masking-cost}, we assume personal mask expenditure of \$0.32 per day or, equivalently, \$2.24 per week, which approximates the cost of one surgical mask per day or one N95 mask per week \citep{skarp-npi-cost-review}.

In the first year of the pandemic, US states steadily ramped up the number of SARS-CoV-2 PCR tests administered each day at a consistent linear pace. Indeed, after running least-squares regression of the cumulative number of tests administered in a state on each day against time (squared) using test data obtained from the COVID Tracking Project \citep{ctp}, we obtain $R^2$ values above 0.97 for all states. Across states, the linear rate of testing capacity increase varies from an additional 7 to 40 tests per million population per day. Our cost function accounts for this by assuming that the number of tests administered in a given week under mask mandate is a linear function of the cumulative number of past weeks spent under mask mandate, with the slope given by the state-specific rate of testing capacity increase obtained from the regression. This yields the total number of tests administered in a state in any given week under the specified masking policy. We convert this quantity to a dollar value assuming that each test costs \$100 based on \citet{skarp-npi-cost-review,lo-test-cost,sharfstein-test-cost}, which includes the cost of procuring the test as well as labor for sample extraction and diagnostic lab testing.

We similarly assume that, while contact tracing policies are in place, tracing capacity ramps up at a linear pace. This is in line with increases over time in capacity reported in wide-scale assessments of US contact tracing programs \citep{lash-tracing,rainisch-tracing}, as well as general increases over time in state-level hiring of contact tracers reported in media \citep{npr-tracing,ama-tracing}. Fitting a log-normal model to data from \citet{lash-tracing}, we estimate the mean number of cases interviewed per week per 100,000 population to be 95.0
during their period of study, June--October 2020. Similarly, based on data from \citet{rainisch-tracing}, we estimate a mean of 170.5 cases interviewed per week per 100,000 population during November 2020--January 2021. With four months separating August and December 2020 (the midpoints of the respective study periods), we therefore assume an increase in capacity of $(170.5 - 95.0)/16 \approx 4.72$ cases interviewed per 100,000 population per week while contact tracing policies are active. We convert this number to a dollar value based on the average cost of contact tracing per index case. \citet{fields-tracing} report the hourly cost of contact tracing at $\$107.22/4.16 \approx \$25.77$. According to \citet{spencer-cdc-tracing}, the median caseload per investigator during their two-week evaluation period was 31. Assuming a 40-hour work week, this implies a cost per case of $\$25.77\times 80/31 \approx \$66.50$. This number, which we take as our cost of contact tracing per index case, is near the midpoint of the interval reported in \citet{skarp-npi-cost-review} (\$40.73--\$93.59) based on different data. Table \ref{tab:health-params} records the testing, tracing, and masking cost parameters with references.


\begin{table}[htbp!]
\centering
\caption{Economic parameters. All costs are in USD2020.}
\begin{tabular}{|p{40mm}|p{40mm}|p{70mm}|}
\toprule
\textbf{Parameter}  & \textbf{Value} & \textbf{Reference} \\
\midrule 2019 GDP per capita by state & 
Varying 
(\$39,000--211,000) 
& \citet{bea-gdp}\\
\hline 2019 per capita income by state & 
Varying 
(\$39,000--85,000) 
& \citet{bea-expenditure}\\
\hline 2019 US median personal income & 
\$35,980
& \citet{census-median-income}\\
\hline 2019 population by state & 
Varying 
(0.575--39.5 million)
& \citet{bea-income}\\
\hline 2019 US GDP current dollar growth rate & 4.1\% & \citet{bea-gdp-growth}\\
\hline Value of a statistical COVID death (VSCD) & 
Low: \$4.47 million \newline
High: \$10.63 million
& \citet{robinson-vsl}
\\
\hline Voluntary social distancing cost per COVID infection by state & Varying (\$1,491--3,199) & \citet{aum-korea} \\
\hline Productivity cost of COVID infection & One week of state median income & 
\citet{bea-expenditure,skarp-npi-cost-review}
\\
\hline Average medical cost of COVID infection & \$3,045 & \citet{bartsch-healthcare-cost,demartino-healthcare-cost} \\
\hline Net present value of GDP loss due to learning loss by state & 
Low: 9\% GDP per 0.33 school-years \newline
High: 69\% GDP per 0.33 school-years
& \citet{psacharopoulos-learning}\newline
\citet{hanushek-learning}\\
\hline Learning loss accrued during COVID & 0.35 school-years & \citet{betthauser-learning-review,fahle-us-learning-loss} \\
\hline Direct GDP loss due to school closure & 0.2\% GDP per four weeks  & \citet{lempel-school-cost} \\
\hline Employment rate decrease due to workplace closure 
& 
Low: 2\%; 
Mid: 4\%; 
High: 6\% 
& \citet{gupta-labor-effects,barrot-business-effect,crucini-orders} \\
\hline Employment rate decrease due to social distancing mandates 
& 
4\%
& \citet{gupta-labor-effects,crucini-orders,baek-unemployment,coibion-macro,bodenstein-social}
\\
\bottomrule
\end{tabular}
\label{tab:econ-params}
\end{table}


\begin{table}[htbp!]
\centering
\caption{Masking, testing, and tracing parameters. All costs are in USD2020.}
\begin{tabular}{|p{40mm}|p{40mm}|p{70mm}|}
\toprule
\textbf{Parameter}  & \textbf{Value} & \textbf{Reference} \\
\midrule Daily personal mask expenditure & \$0.32 & \citet{bartsch-masking-cost,skarp-npi-cost-review}\\
\hline Cost of a PCR test & \$100 & 
\citet{skarp-npi-cost-review,lo-test-cost,sharfstein-test-cost}\\
\hline Daily rate of testing capacity increase by state & Varying (7--40 tests per million pop.) & \citet{ctp} \\ 
\hline Cost of contact tracing per index case & \$66.50 &
\citet{skarp-npi-cost-review,fields-tracing,spencer-cdc-tracing}\\
\hline Weekly rate of contact tracing capacity increase & 4.72 cases per 100k pop. & 
\citet{lash-tracing,rainisch-tracing}
\\ 
\bottomrule
\end{tabular}
\label{tab:health-params}
\end{table}

\section{Results}
\label{sec:results}

\subsection{Epidemiological model}

\begin{figure}[t!]
    \centering
    \includegraphics[width=\textwidth]{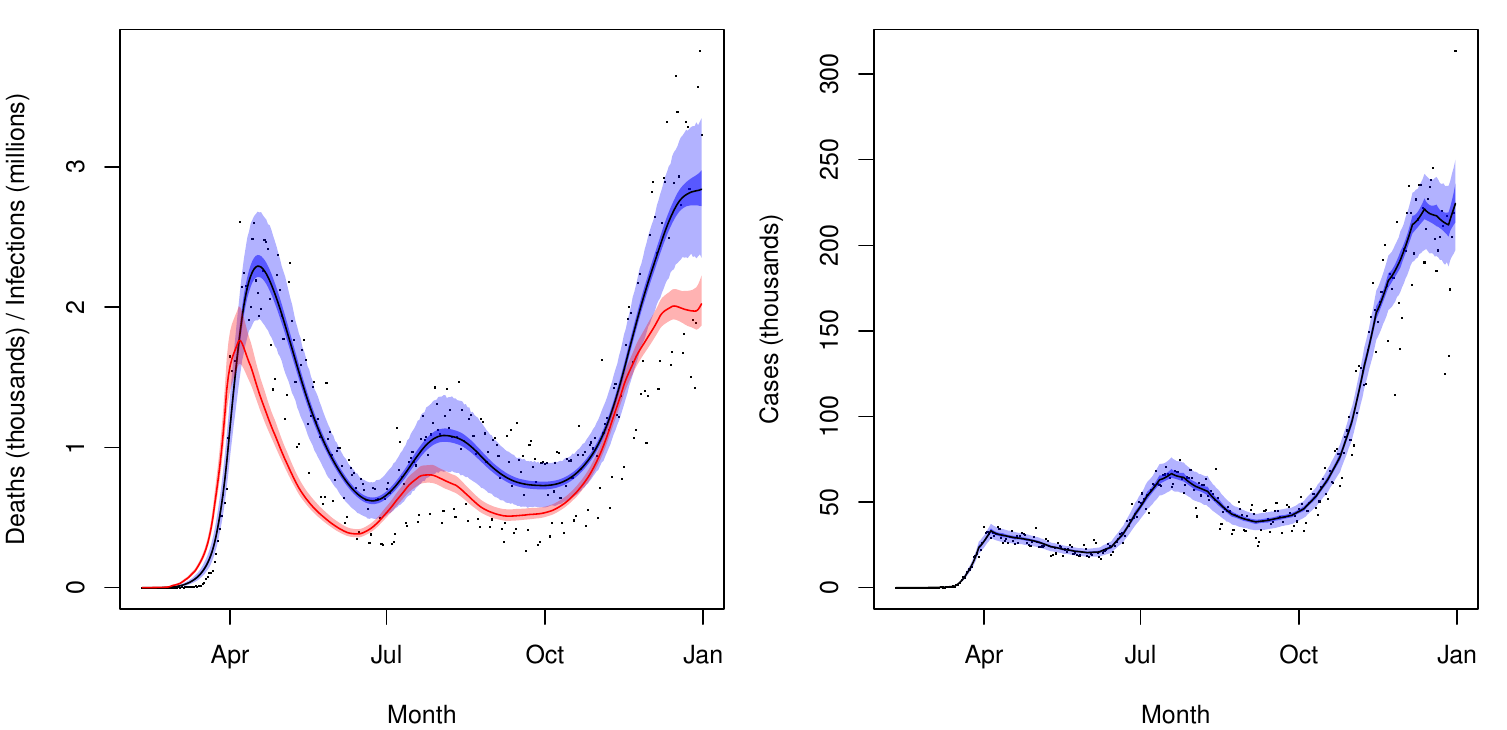}
    \caption{State-specific SEIRD results aggregated to the US. Observed deaths $d(t)$ and cases $c(t)$ are plotted in black. Posterior median and 90\% credible intervals of the underlying mean parameters $m_D(t)$ and $m_C(t)$ are in dark blue. 90\% credible intervals of the posterior predictive distributions of $d(t)$ and $c(t)$ are in light blue. On the left, posterior median and 90\% credible intervals for active viral prevalence $I(t)$ are in red.}
    \label{fig:us}
\end{figure}

Figure \ref{fig:us} displays estimates of active viral prevalence and posterior predictive distributions of the observed deaths and cases in the US in 2020. While the posterior predictive distributions of deaths are well-calibrated at the state-level (as evidenced by Figure \ref{fig:alaska}, for example), when aggregated to the US as a whole they exhibit under-coverage. This is because we model the states independently and do not explicitly account for the ``weekend effect'', i.e., consistent under-reporting of deaths on weekends which leads to highly correlated residuals across states on those days. As we evaluate and optimize NPI policies at the state-level, however, this does not pose an issue for our downstream analysis.
See \citet{irons} for more detailed reports and discussion of state-specific prevalence estimates.

We estimate that there were 58.5 (95\% CI: 55.7--62.7) million COVID infections in the US in 2020, representing about 18\% of the population. 
Weighting the posterior state-level IFR ($\iota_s$) estimates by the proportion of 2020 U.S. COVID deaths occurring in each state, we obtain a national IFR of 0.78\% (0.74\%--0.82\%).
Our findings are on par with the systematic meta-analysis of \citet{ifr-meta}, who estimated an IFR of 0.68\% (0.53\%--0.82\%) for COVID in 2020 based on 24 studies from a range of countries, as well as \citet{ifr-england}, who estimated the IFR in England in 2020 based on a series of nationally representative testing surveys at 0.67\% (0.65\%--0.70\%). Similarly, \citet{ward-ifr-england} estimated the IFR in England in October 2020 at 0.74\% (0.48\%--1.40\%).

Our estimates of SARS-CoV-2 infections incident in 2020 leverage prior work based on random sample testing \citep{irons}---a putatively unbiased measure of viral prevalence---and, as noted above, produce an IFR similar to that estimated in England in 2020 also based on representative testing surveys. Nevertheless, we note that our findings concerning policy evaluation and optimization below are robust to sensible variations in the IFR. This is because the costs of infections are dominated by COVID deaths, which are identified from the clinical data we use here and, therefore, are outputs of our model not substantially affected by the IFR parameter (which only varies the estimated number of infections incident per death).

\subsection{NPI regression model}

For brevity, here we mainly report results for the robust log-linear regression model (ii) defined in Section \ref{sec:model:regression}, which controls for deaths and removals, but not infections, incident in the prior week. 
Qualitative conclusions are practically identical across models, with the main distinction being that controlling for more terms tends to attenuate the effect of school closure in reducing transmission rates. 

\subsubsection{Fit to data}
\label{sec:results-fit}

Regarding the fit to data, the posterior median $R^2$ 
is 0.59,
indicating that a substantial proportion of the variance in transmission rates remains unexplained by NPIs or behavioral response to the fear of infections. Indeed, there are numerous factors affecting SARS-CoV-2 transmission---including super-spreader events and introduction of new infections from outside the state---accounted for by the error terms $\varepsilon^{(s)}(w)$ that are difficult to predict. Figure \ref{fig:rt} plots the \emph{maximum a posteriori} (MAP) trajectory of $R_0$ output by the epidemiological model in the four most populous states---California, Texas, Florida, and New York---against the posterior predictive distribution from the NPI model fit to this output. The model fits the data well, but cannot capture unpredictable shocks in transmission, which are reflected in future predicted values of the transmission rate through the AR(1) term. 
The 
AR(1) parameter $\varphi$ 
is 0.76 (0.68--0.83),
indicating a high degree of residual autocorrelation in the weekly-varying reproduction number $R_0^{(s)}(w)$ across states. 
Finally, the posterior degrees of freedom $\nu_\varepsilon$ for the Student-$t$ distributed residuals $\varepsilon^{(s)}(w)$ is 2.9 (2.3--3.7), indicating that a heavy-tailed error distribution is appropriate for these data.

\subsubsection{Total effect of NPIs}

\begin{figure}[t!]
    \centering
    \includegraphics[width=\textwidth]{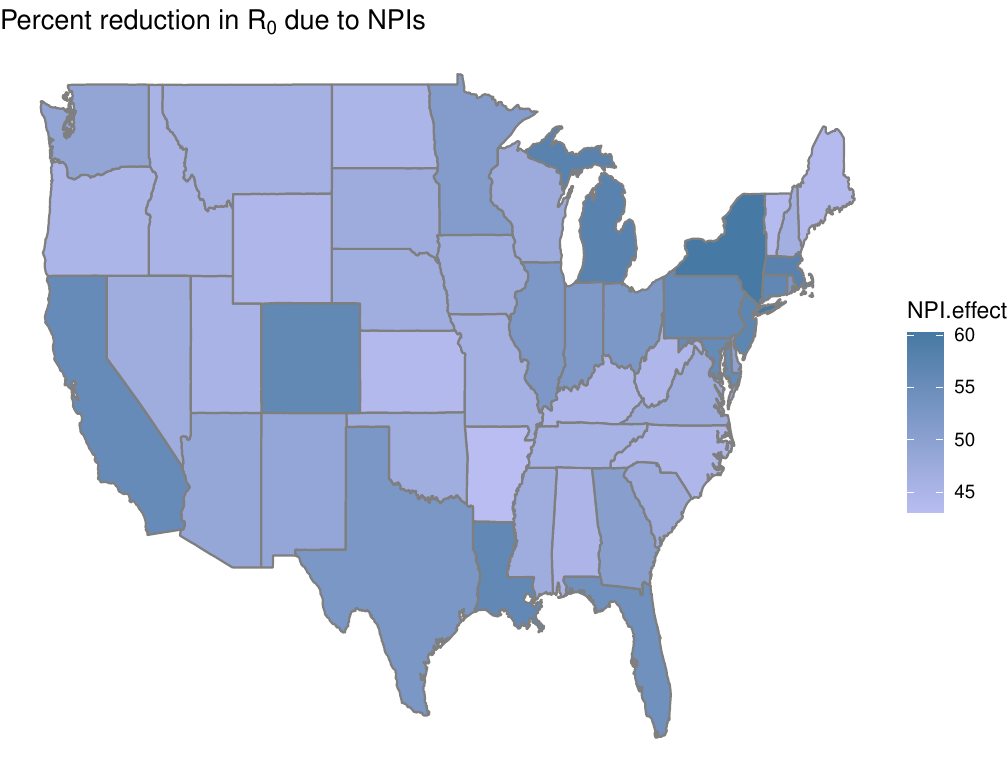}
    \caption{Posterior median total percent reduction in $R_0$ due to NPIs by state.}
    \label{fig:npi-map}
\end{figure}

We estimate a pooled baseline reproduction number $R_0$ under no interventions of 
2.3 (95\% CI: 2.0--2.7). 
This is on par with the systematic review of \citet{liu-r0}, who report $R_0$ estimates for wild-type SARS-CoV-2 with median 2.79 and interquartile range 1.16. 
The pooled total effect of NPIs, $\alpha=\sum_{k=1}^p \beta_u(k)$, which represents the effect of ``full lockdown''---i.e., $u_k(w)=1$ for all $k=1,\ldots,p$---yields a reduction of $R_0$ by 
by 50.0\% (38.5\%--59.3\%) to 1.14 (1.03--1.28).
By comparison, the pooled effect of deaths per 1,000 population, $\beta_D$, 
is -0.14 (-0.41--0.13) and the pooled effect of removals per 10 population, $\beta_R$, 
is -0.41 (-0.91-- -0.03).
When evaluated at the mean weekly rates of deaths and removals across states in 2020, this yields a combined
11.2\% (6.5\%--16.7\%) 
reduction in $R_0$---or 22\% (12\%--39\%)
of the total effect of NPIs---from voluntary social distancing and other protective measures due to fear of infection.

For a population of this kind, NPIs alone would most likely not be sufficient to suppress transmission at the start of the outbreak ($R_0 < 1$). However, voluntary 
protective measures 
and acquired immunity in combination with full lockdown would be enough to effectively suppress viral spread (at least in the absence of exogenous shocks). Figure \ref{fig:counterfactual-deaths} exhibits posterior trajectories of deaths in the US under various NPI strategies. 
Under full lockdown in 2020, cumulative deaths would have been 98,850 (41,531--201,532), about one quarter of the 348,949 actually observed.

Our estimate of the total percent reduction in $R_0$ due to NPIs is more conservative than others reported in the literature. \citet{flaxman-effects}, \citet{brauner-effects}, and \citet{banholzer-npi-effect}, respectively, find 81\% (75\%--87\%), 77\% (67\%--85\%), and 67\% (64\%--71\%) reductions in transmission in the initial spring 2020 wave. Studying the second wave, \citet{sharma-effects} report a combined NPI effect of 66\% (61\%--69\%). We note that none of these studies control for confounding (e.g., endogenous social distancing), which may account for the discrepancy with our estimates. Indeed, when we add the effect of deaths and removals incident in the prior week to that of NPIs, the combined reduction in $R_0$ approaches these higher estimates. Another possible explanation is the context: we study the US whereas \citet{flaxman-effects,brauner-effects,banholzer-npi-effect,sharma-effects} focus primarily on European countries, which may have implemented stricter NPIs or practiced greater adherence to restrictions, and which exhibited higher $R_0$ values (3.3--3.8), perhaps due to earlier introduction of the virus to European countries or higher levels of social mixing on average.

Zooming in on the state-level results, we can similarly quantify the total effect of NPIs on transmission in state $s$ by 
\begin{linenomath*}
\[
\alpha^{(s)} := \sum_{k=1}^p \beta^{(s)}_u(k),
\]
\end{linenomath*}
with $p^{(s)}=100(1-\exp(\alpha^{(s)}))$ representing the total percent reduction in $R_0^{(s)}$ under full lockdown. Figure \ref{fig:npi-map} displays the geographic distribution of the posterior median of $p^{(s)}$ across states.
Overall, NPIs tend to be more effective in more urbanized and populous states. Some of this variation may be explained by the literature on political polarization and partisan social distancing during the pandemic \citep{allcott-polarization,barrios-politics,painter-politics,brodeur-review,adolph-politics-masks,adolph-politics-policies,adolph-politics-timing}. Alternatively, and related to our discussion in the previous paragraph, we note that rural states tend to have lower baseline $R_0^{(s)}$ values, possibly due to later importation of the virus and lower levels of social mixing. With a lower ceiling in these states, NPIs have less room to suppress transmission.

Finally, note that the $p$-vector
\begin{linenomath*}
\[
\rho^{(s)} := \beta^{(s)}_u/\alpha^{(s)}
\]
\end{linenomath*}
consists of weights representing the proportional contribution of each NPI to the total reduction of transmission. Here $\rho^{(s)}$ can be thought of as defining a data-driven ``stringency index'' combining NPIs according to their strengths in a single-number summary of the stringency of government restrictions, as opposed to previously defined measures, such as OxCGRT's stringency index, which average NPIs uniformly without regard for their varying effects on transmission \citep{oxcgrt}.

\subsubsection{Effects of individual NPIs}

\begin{figure}[t!]
    \centering
    \includegraphics[width=\textwidth]{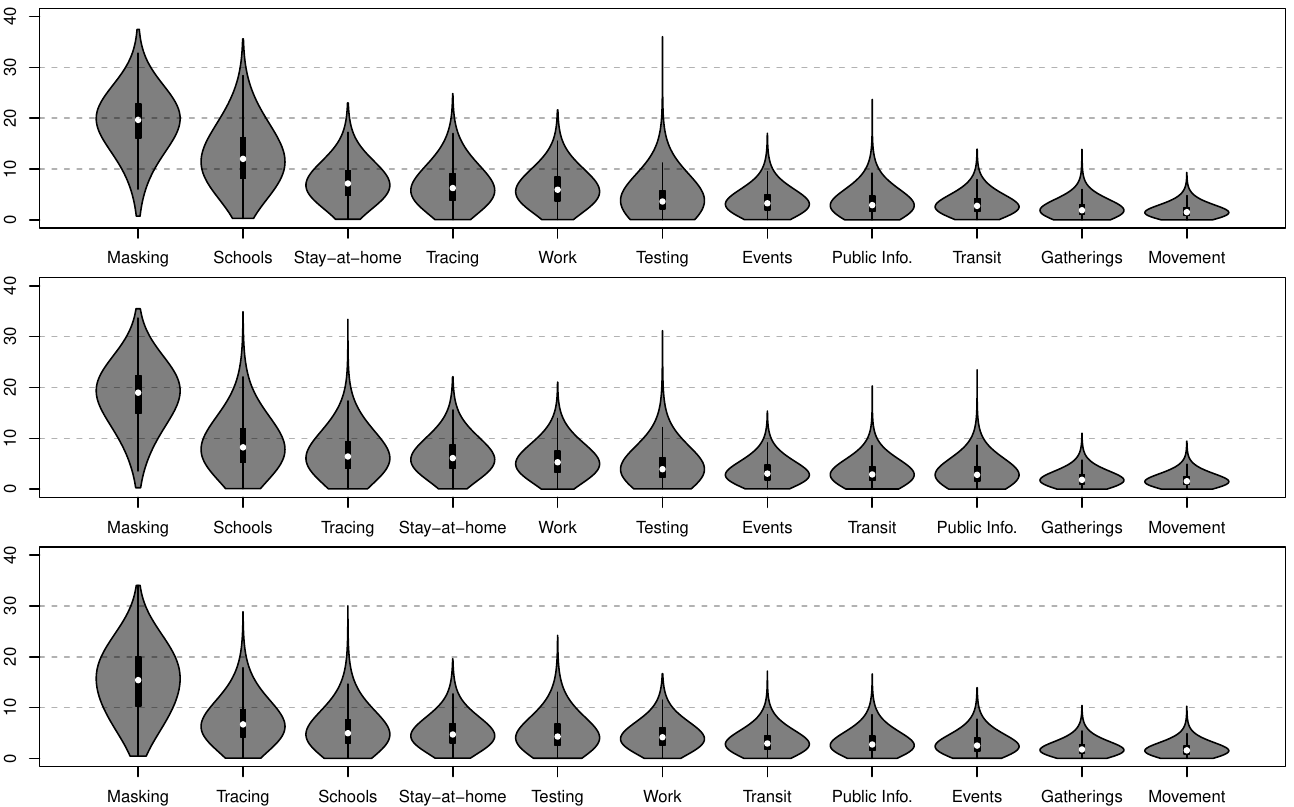}
    \caption{Posterior violin plots of global NPI effects, quantified as the percent reduction in $R_0$, across regression models (i)--(iii) from top to bottom, respectively.}
    \label{fig:npi-effects}
\end{figure}

Figure \ref{fig:npi-effects} shows the pooled effect of each NPI, $\beta_u(k),k=1,\ldots,p$, quantified as a percent reduction in $R_0$, across the robust log-linear regression models defined in Section \ref{sec:model:regression}. 
As above, we focus discussion on model (ii).
Mask mandates are the most effective intervention, reducing $R_0$ by 
19.0\% (6.1\%--28.5\%). 
As shown in Figure \ref{fig:npi-effects}, this effect attenuates from 19.7\% (7.3\%--28.5\%) to 15.4\% (3.3\%--26.9\%) as we move from model (i) to model (iii), controlling for removals and infections incident in the prior week.
By comparison, \citet{sharma-effects} estimate that mask mandates reduced transmission rates by 12\% (7\%--17\%) in the second wave in Europe. 
Based on data from 190 countries between January and April 2020, \citet{bo-npi-effects} conclude that mask mandates were associated with a 15.1\% (7.9\%--21.8\%) decline in transmission.
\citet{KARAIVANOV-mask-effect} find a 22 percent weekly reduction in new COVID-19 cases due to mask mandates in Canada in the summer of 2020. 
Studying the 2020 spring wave in New York City, \citet{yang-npi-effect} find that masking was associated with a 7\% transmission reduction overall and up to 20\% reduction for people over age 65.
Estimating the causal effects of a number of interventions in the US, \citet{chernozhukov-masks} demonstrate that masking policies were highly effective, leading to a reduction in the weekly growth rate of cases and deaths by more than 10 percentage points, with their conclusions holding robustly across model specifications; on the other hand, the effects of stay-at-home orders and business and school closures are much more uncertain. Qualitatively, our results are consistent with a number of other studies demonstrating the efficacy of mask mandates and the protective effects of face mask use \citep{jamison-behavior,talic-mask-effect,li-mask-effect,lyu-mask-effect,rader-mask-effect,greenhalgh-masks}.


Behind mask mandates, in our model (ii) school closure 
reduces the transmission rate by 
8.2\% (1.5\%--20.2\%). 
As shown in Figure \ref{fig:npi-effects}, this effect attenuates from 12.0\% (3.0\%--24.8\%) to 5.0\% (0.8\%--14.3\%) as we move from model (i) to model (iii), controlling for removals and infections incident in the prior week.
By comparison, \citet{brauner-effects} and \citet{banholzer-npi-effect} find that school closures led to 38\% (16\%--54\%) and 17\% (-2\%--36\%) transmission reductions in the first wave of 2020, respectively, and \citet{sharma-effects} estimate a 7\% (4\%--10\%) transmission reduction due to school closures in the second wave. Studying influenza outbreaks, \citet{cauchemez-school-sentinel} found that school holidays led to a 20–29\% transmission reduction among children with no detectable effect on transmission among adults.
Qualitatively, our results are consistent with a number of other studies finding school closures to be one of the NPIs most effective in reducing transmission \citep{markel-spanish-flu-npis,auger-schools,liu-npi-effects,stokes-npi-effects,li-npi-effects,haug-npi-effects,ferguson-strategies}.


As with school closures, we are unable to rule out small effects for the remaining NPIs. Workplace closure reduced transmission by 
5.3\% (0.9\%--12.5\%).
By comparison, \citet{brauner-effects}, \citet{sharma-effects}, and \citet{banholzer-npi-effect} find that business closures led to a 27\% (-3\%--49\%), 35\% (29\%--41\%), and 18\% (-4\%--40\%) reduction in $R_0$, respectively.
The combination of social distancing measures\footnote{As in Section \ref{sec:cost-social}, we define social distancing measures as the combination of stay-at-home orders, restrictions on gatherings, restrictions on internal movement, public information campaigns, public transit closures, and public event cancellations.} yields a 
18.9\% (10.9\%--28.1\%)
reduction in $R_0$, which is on par with the individual effects of mask mandates. 
Looking at individual social distancing measures, we estimate that stay-at-home orders reduced $R_0$ by
6.1\% (1.2\%--14.1\%), 
which is comparable to other estimates in the literature: \citet{bodenstein-social} 
report a 6.5\% transmission reduction; \citet{brauner-effects} report a 13\% (-5\%--31\%) reduction; and \citet{banholzer-npi-effect} report a 4\% (-6\%--17\%) reduction. 
Restrictions on gatherings reduced $R_0$ by 
a modest 
1.9\% (0.3\%--5.7\%).
However, a number of other studies estimate large effects of strict gathering restrictions: \citet{brauner-effects} report a 42\% (17\%--60\%) transmission reduction; \citet{sharma-effects} report a 26\% (18\%--32\%) reduction; \citet{banholzer-npi-effect} report a 37\% (21\%--50\%) reduction; 
and \citet{bo-npi-effects} report a 42.9\% (41.6\%--44.2\%) reduction associated to social distancing measures more broadly.\footnote{We note that these are association studies based on observational data, i.e., they do not control for potential confounders.}
Nevertheless, even with our relatively conservative estimates of the effects of social distancing measures, we find that they are cost-effective interventions in combination, as we demonstrate in Section \ref{sec:oc-results}.

Finally, our estimates of the effects of testing and tracing policies---which yield 
3.9\% (0.6\%--12.3\%) 
and 
6.4\% (1.1\%--16.0\%) 
reductions in $R_0$, respectively---allow for the possibility of both small and large effects on transmission. Evidence on the effectiveness of contact tracing, in particular, is mixed. \citet{rainisch-tracing} concluded that case investigation and contact tracing were effective in reducing transmission based on their estimates of the number of COVID-19 cases and hospitalizations averted by these measures in the US. \citet{wang-tracing} find that testing and case isolation were effective, but the effect of contact tracing is marginal due to slow follow-up times in case investigation. They note that contact tracing can be more effective if follow-up is accelerated. In modeling studies, \citet{hellewell-tracing} and \citet{davis-tracing} find that contact tracing can be effective if carried out well---with the latter reporting a potential 15\% reduction in $R_0$---but that its effectiveness is dependent on a number of epidemiological and implementation-related factors, including tracing coverage and speed.
Nevertheless, even allowing for small effects, we find that testing and tracing are highly cost-effective NPIs owing to their low cost relative to other interventions.

\subsection{Evaluating and optimizing costs}
\label{sec:oc-results}


\begin{figure}[htbp!]
    \centering
    \includegraphics[width=\textwidth]{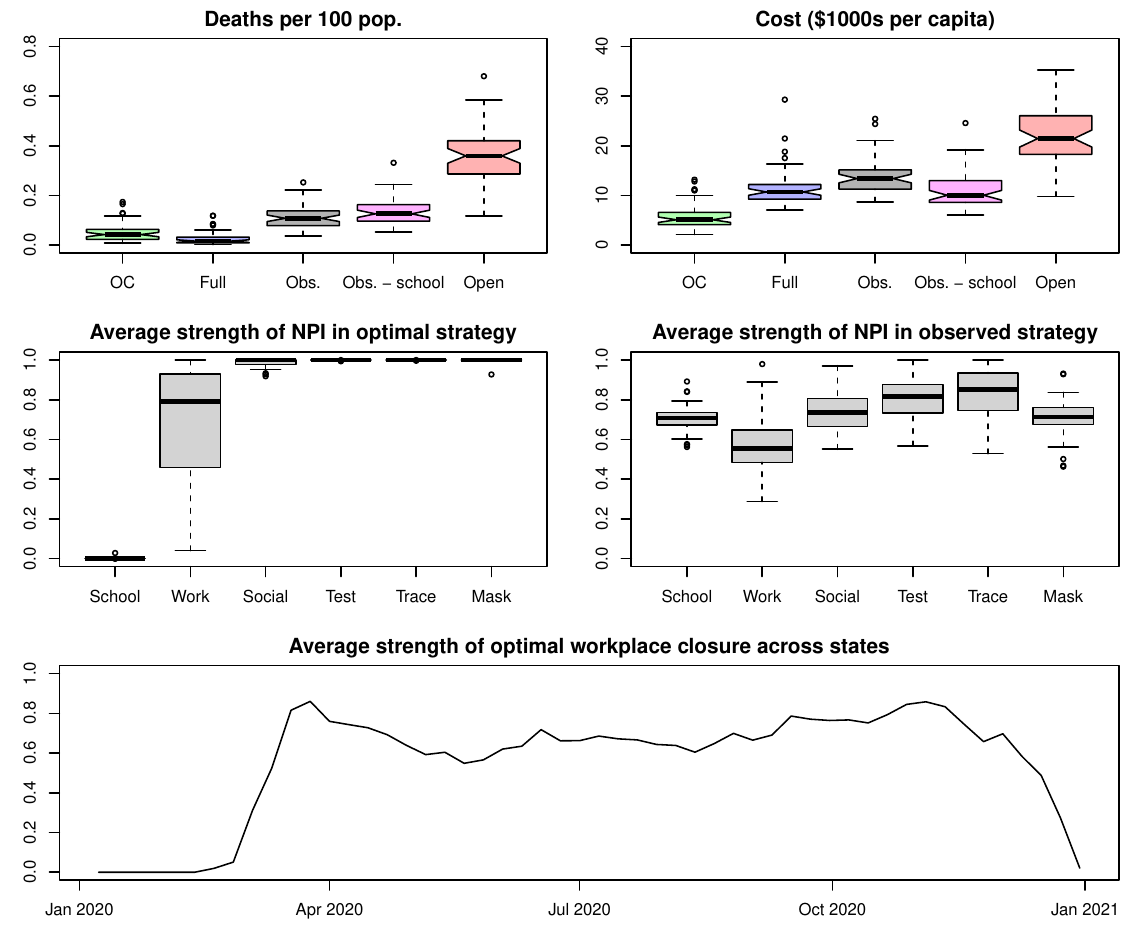}
    \caption{
    \textbf{Top panels:} Boxplots of the following quantities across states: posterior median of deaths per 100 population incurred by the optimal control (OC), full lockdown (Full), observed (Obs.), observed minus school closures (Obs. - school), and fully open (Open) policies; expected total cost in thousands of USD2020 per capita incurred by each policy.
    \textbf{Middle panels:} the average strength of each NPI in the optimal and observed strategies.
    \textbf{Bottom panel:} the average strength of optimal workplace closures across states in each week.
    }
    \label{fig:oc-boxplots}
\end{figure}

\begin{figure}[t!]
    \centering
    \includegraphics[width=\textwidth]{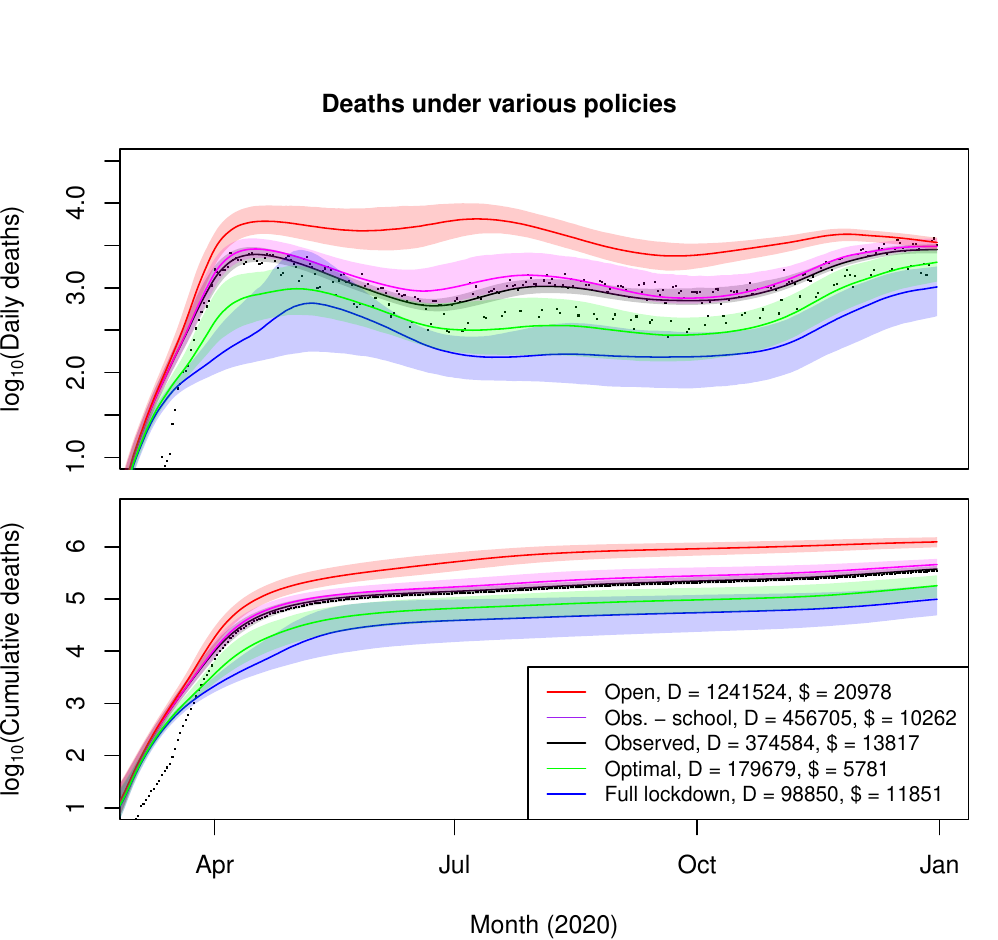}
    \caption{Posterior trajectories of daily deaths and cumulative deaths in the US under various strategies. The legend records the posterior median of the total deaths and expected total cost per capita in USD2020 incurred by each policy.}
    \label{fig:counterfactual-deaths}
\end{figure}

\paragraph{Baseline scenario.}
\label{sec:results-baseline}
Figure \ref{fig:oc-boxplots} displays the results of our policy evaluation and optimization methodology detailed in Section \ref{sec:methods-cost} under the baseline cost scenario, which uses a cost function based on the medium value of the cost of workplace closure, the low VSCD (i.e., adjusted for the age pattern of COVID deaths), and the low cost of learning loss (i.e., assuming distance learning is 90\% as effective as in-person schooling) listed in Table \ref{tab:econ-params}. The upper panels of Figure \ref{fig:oc-boxplots} display boxplots of the COVID death rate and per capita cost (including life costs) incurred by various policies across states. We consider the following five policy strategies: optimal control (OC), which is the strategy optimizing the cost function \eqref{eq:cost}; full lockdown (Full), which assumes all NPIs are enforced at their strictest level for the entire year; observed (Obs.), the policy actually implemented; the observed policy minus school closure (Obs. - school); and the open policy (Open), which assumes no use of NPIs. 

Full incurs the fewest deaths---as we would expect---followed by OC, then Obs., then Obs. - school, and finally Open. Regarding overall costs, OC is the least expensive policy (by definition), followed by Full, then Obs. and Obs. - school (which are approximately on par), and finally Open. The middle panel of Figure \ref{fig:oc-boxplots} displays boxplots of the average strength of each NPI in the optimal control and observed strategies over the year across states. An average strength of 1 implies that the intervention is implemented in its strictest sense for the entire year uninterrupted; an average strength of 0 implies that the intervention is never implemented at all. 

The OC strategies in each state are uniformly comprised of: consistent and strict use of social distancing measures, testing, tracing, and masking; no use of school closure; and moderate to strong use of workplace closure. The bottom panel of Figure \ref{fig:oc-boxplots} displays the average strength of workplace closure across states over time. Generally, the optimal strategy involves ramping up workplace closure to combat new waves of infections, with implementation peaking in response to the spring, summer, and fall waves of 2020. In Section \ref{sec:sensitivity} we explore the sensitivity of these results across models and plausible perturbations of the cost function. Our qualitative findings are robust across the range of scenarios.

Figure \ref{fig:counterfactual-deaths} displays the result of aggregating the state-level policy outcomes to the US as a whole. 
The open strategy incurs the most deaths and is the most expensive by far, with its cost arising entirely from infections. We estimate that 
1.24 (0.94--1.58)
million COVID deaths would have occurred in 2020 in the absence of public health interventions---on par with but lower than the 2.2 million projected by \citet{ferguson-npi-impacts}, who do not account for the endogenous social response to the virus. The burden of infections under the open strategy yields an expected cost per capita of 
\$20,978
USD2020, which translates to a gross impact of 
\$6.9 trillion---about 
32\%
of US GDP in 2019 \citep{bea-gdp-growth}. On the other hand, under full lockdown, we would have observed only
98,850 (41,531--201,532)
COVID deaths and a cost to society of 
\$11,851
per capita, which are surpassed by the 
374,584 (337,313--445,227)
deaths and
\$13,817
per capita ( 
\$4.6 
trillion total, or 21\% of 2019 GDP) lost under the observed policy. 
Note, however, that full lockdown can become equally or more expensive than the observed and fully open policies if we assume a high cost of learning loss. See Section \ref{sec:sensitivity}. 
The posterior median mortality rate under full lockdown is
$98,850/329.5\approx 300$ 
deaths per million, which is about 80\% of the COVID mortality rate observed in Canada in 2020 \citep{owidcoronavirus}.

The observed policy minus school closure (Obs. - school) incurs 
456,705 (370,951--626,540)
deaths,
which is larger than the 
374,584 (337,313--445,227) 
deaths under Obs.
However, the expected per capita cost to society of Obs. - school (including life costs), \$10,262, is lower than that of Obs., \$13,817. 
To some degree, conditional on the other NPIs implemented, the decision to close schools presented a marginal trade-off between the learning of students and the health of those most vulnerable to COVID. The extended school closures that were enacted across the country in 2020 prioritized the latter in lieu of the former. While our model estimates that they saved approximately 
77,168 (12,268--235,954)
lives, this came at the expense of \$2 trillion in lost learning, or
\$25.9 (8.4--156.4) million per COVID death.
However, this trade-off was not inevitable. Under the optimal policy (OC), which involves no closure of schools in 2020 beyond the usual 16 weeks of break, we incur 
179,679 (95,698--309,863)
deaths,  195,481 (67,390--300,539) fewer than under Obs.,
 at an expected cost to society of only
\$5,781. 
Most of the savings relative to the observed policy stem from the cost of infections and school closure. Hence, more timely, stringent, and enduring use of other measures---social distancing, testing, tracing, and masking, along with reactive workplace closures---would have been sufficient to limit COVID mortality substantially below what was observed without incurring profound learning loss.

We can compare our estimates of the gross impacts of COVID-19 and NPIs to others in the literature. 
\citet{summers,ifp-cost-pandemic} project the total cost of the pandemic in the US over its full duration at \$16 trillion, which is about 3.5 times our estimate of \$4.6 trillion in losses observed during 2020.
\citet{flaxman-effects} estimated that NPIs averted 3.1 (2.8–3.5) million deaths up to May 2020 in 11 countries totaling 375 million population. 
\citet{ferguson-npi-impacts} predicted that NPIs would prevent 1.1 million deaths in the US over the course of the pandemic.
\citet{greenstone-distancing} projected that moderate social distancing would save 1.7 million lives in the US by October 1, with mortality benefits of \$8 trillion, or \$24,000 per capita, based on the same VSCD used in our baseline scenario. Highlighting the inter-generational transfer of wealth stemming from the implementation of social distancing measures, they note that the vast majority of the monetized benefits of social distancing accrue to people age 50 or older.
More conservatively, \citet{eichenbaum-macro} find that containment policies, if implemented optimally, would save about half a million lives in the US based on low values of key epidemiological parameters---specifically, they use an IFR of 0.5\% and an $R_0$ of 1.45 for their baseline model.
\citet{thunstrom-costs-benefits} estimated that social distancing in the US would save 1.24 million lives at a cost of \$7.2 trillion in lost GDP, which implies that social distancing measures would yield net losses for any VSCD below $7.2/1.24\approx 5.8$ million dollars. To the contrary, we find that NPIs (and social distancing measures in particular) are cost-effective for a VSCD of \$4.5 million.
Assuming a vaccine arrives (stochastically) after a year to end the pandemic, \citet{farboodi2021} estimate the per capita cost of the optimal policy at \$8,100, comparable to our \$5,781. They find that the \emph{laissez-faire} equilibrium (i.e., in the absence of government intervention) would only incur a cost of \$12,700 per person, as compared to our \$20,978.
Undertaking a cost-benefit analysis of confinement policies targeted toward mitigation and (strict) suppression, \citet{gollier-cost} finds that both strategies incur a total cost---combining economic and life costs---equating to 15\% of annual GDP, or about \$10,000 per capita, which is comparable to our estimates of the total costs of the various containment strategies in Figure \ref{fig:counterfactual-deaths}. \citet{jones-oc} estimate that, under an optimal social distancing policy, GDP declines by 12\% and 0.17\% of the population (about 560,000) die from COVID in the first 26 weeks of the pandemic. 
In a simulation study, \citet{keogh-brown-covid-cost} estimate that containment strategies in the UK to suppress COVID-19 through the end of 2020 would incur health costs of 1.7\% of GDP and economic costs of 29.2\% of GDP, with 7.3 percentage points coming from workplace absenteeism of parents affected by school closure and 21.9 percentage points from business closure.
Their total cost is comparable to our estimate 
for the U.S. in 2020 (i.e., under the observed policy), which is 21\% of GDP. However, relative to the results of \citet{keogh-brown-covid-cost}, we find that health impacts are a much larger portion of the total, costs related to business closure are much smaller, and costs related to school closure are somewhat larger.

\paragraph{Sensitivity analysis.}
In Section \ref{sec:sensitivity}, we discuss sensitivity analysis of our results over a range of NPI regression model and cost function specifications. 
Our qualitative findings about the structure of optimal policies are robust across scenarios, with the main quantitative distinction being the optimal strength of workplace closure. We also find that the relative ranking by cost of the policies considered can vary across cost function specifications.

\section{Discussion}
\label{sec:discussion}

We have developed a statistical decision framework in order to conduct a cost-effectiveness analysis of non-pharmaceutical interventions in the U.S. during COVID-19. 
We note that, for practical purposes and due to lack of available data, our model of SARS-CoV-2 transmission does not account for a number of complexities. 

We do not explicitly account for the age structure of a state's population and its infections, 
although these are reflected in the state-specific IFR estimates used in our model \citep{irons}. As such, our reported effects of NPIs on transmission and costs associated to infections and NPIs should be interpreted as aggregate measures.

We do not model state-level hospital capacity and potential excess costs or deaths arising from an overwhelmed medical system.
In principle, doing so would serve to increase the cost associated to infections. However, estimates of the IFR in England, which experienced COVID death rates similar to the U.S. in 2020 \citep{owidcoronavirus}, are fairly constant over the year \citep{ifr-england}, suggesting that COVID mortality outcomes---the dominant term in the cost of infection---were not highly sensitive to fluctuations in the burden on hospitals.  

We do not account for mental health costs arising from lockdowns and from the fear of infection, which may be substantial \citep{summers,ifp-cost-pandemic}. While there is some recent work estimating the causal effects of stay-at-home orders and school closure on mental health outcomes \citep{ferwana-mental}, the effects of other relevant exposures, including workplace closure, other NPIs, and the pandemic itself, have not been ascertained.

Given the highly correlated implementation of NPIs and that we are already accounting for spatial heterogeneity of their effects in our hierarchical model, we do not also model temporal variation in NPI effects (e.g., arising from ``pandemic fatigue''), which has been documented in some studies \citep{petherick-fatigue,sharma-effects,ge-npi-effects-time-varying}, 
as it would be difficult to identify from the data. 
While \citet{ge-npi-effects-time-varying} find that the overall effect of NPIs in Europe increased over time in 2020, \citet{petherick-fatigue} observe increasing use of masks but declining adherence to physical distancing measures across countries over the year. \citet{sharma-effects} estimate a reduced effect of school closure in the second wave in Europe. In relation to our results, increasing the efficacy of masking policies and decreasing the efficacy of school closures would not reverse the conclusion that mask mandates are highly cost-effective whereas school closures are not. However, substantial decreases in the efficacy of workplace closure and social distancing measures could impact their cost-effectiveness.

Finally, as we model viral spread in the U.S. states independently of each other, we do not account for spillover effects of intervention policies between states, which may play a role in overall trends in transmission \citep{holtz-spillover-effects}.

While the use of NPIs involves health, economic, and social trade-offs, it is not a zero-sum game. As others have noted 
\citep{kaplan-covid-political-economy}, 
appropriately implemented restrictions can simultaneously limit deaths and the aggregate costs to society incurred during a pandemic.
Our methodology enables us to derive optimal NPI strategies, which 
consist of
timely, enduring, and stringent use of testing, tracing, and masking policies, social distancing measures, and reactive workplace closure, with no closure of schools. 

This last finding is salient as schools were closed for extended periods in the U.S. and in many other countries throughout the pandemic. Growing evidence suggests that the impacts on school children are substantial and long-term \citep{unicef-education-crisis,nyt-schools}. 
As we show in Section \ref{sec:sensitivity}, our conclusion that school closures were not cost-effective is robust to plausible variation in the cost function. Furthermore, given that we estimate school closure to be one of the interventions most effective in reducing transmission, our results would not be easily reversed based on different modeling assumptions. If we have overestimated the effect of school closure on transmission reduction for most of 2020, getting closer to the truth would only strengthen our findings. Relative to the first COVID wave in Europe \citep{brauner-effects}, \citet{sharma-effects} found that school closure was substantially less effective in reducing transmission, with their point estimate of the effect (7\%), slightly more than half of our model (ii) estimate (12\%).
\citet{sharma-effects} speculate that the effect attenuated from the first to the second wave because many schools in Europe reopened without substantial increases in transmission. Some have argued that, with adequate health 
protocols in place, 
U.S. schools that remained closed through the 2020--2021 academic year could have resumed in-person learning safely \citep{nyt-schools,viner-schools-reopening}. 

While our study focuses on COVID-19 in the U.S. prior to the arrival of vaccines, our qualitative findings shed light on NPI implementation in other settings. 
Masking, testing, and tracing are relatively cheap and likely to remain cost-effective universally: for severe and relatively mild pandemics; in lower resource settings; and after effective pharmaceutical interventions become available. After the arrival of vaccines and antiviral treatments, workplace closures and social distancing measures should be enacted more sparingly. Although school closures were not cost-effective, evidence suggests that distance learning helped to mitigate learning loss. Consequently, extended school closures are likely to be relatively more costly in low- and middle-income countries with younger populations and less capacity to provide effective education remotely \citep{unicef-education-crisis}. Likewise, with fewer opportunities for remote work and less online economic activity, workplace closures, stay-at-home orders, and other social distancing measures may be more costly in these countries \citep{decerf-lives-livelihoods}. For less virulent diseases with a similar age pattern of death, extended school closures are unlikely to be justifiable, and extended workplace closures and social distancing measures should be mandated with care. If possible, more targeted interventions should be used \citep{acemoglu-targeted}. 

\paragraph{Acknowledgments:}
This research was supported by NICHD grant number R01 HD070936, a Shanahan Endowment Fellowship, a NICHD training grant, T32 HD101442-01, to the Center for Studies in Demography and Ecology at the University of Washington.
N.J.I. was also supported by the Florence Nightingale Bicentenary Fellowship in Computational Statistics and Machine Learning from the University of Oxford Department of Statistics and the Leverhulme Centre for Demographic Science and the Leverhulme Trust (Grant RC-2018-003).

\paragraph{Author contributions:}
N.J.I. and A.E.R. designed the study, conducted the data analysis, and wrote the manuscript.


\printbibliography


\appendix

\section{Sensitivity analysis of optimal control strategies}
\label{sec:sensitivity}

\begin{figure}[t!]
    \centering
    \includegraphics[width=\textwidth]{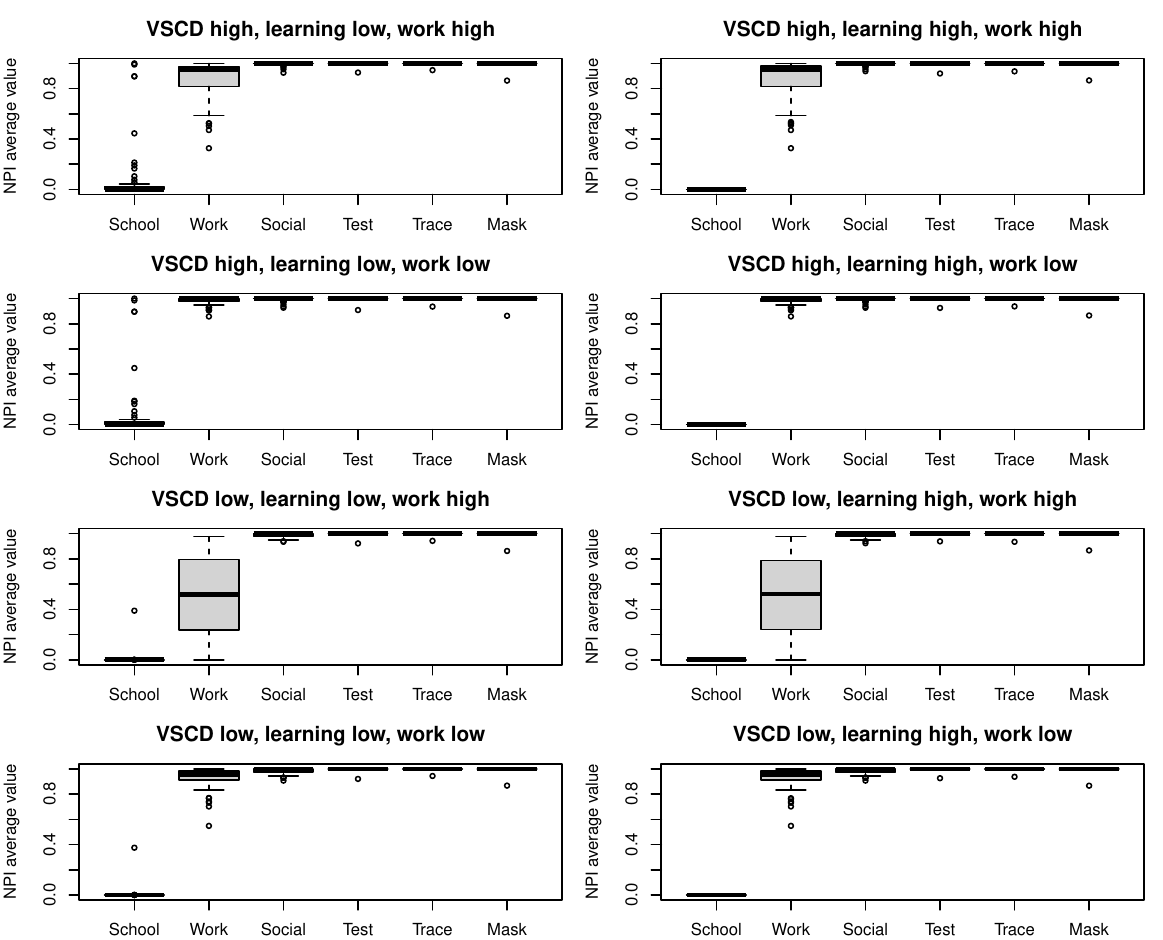}
    \caption{Sensitivity analysis for the optimal control results under regression model (i). Boxplots of the average value of the optimal NPI policy over the year in each state.}
    \label{fig:sensitivity1}
\end{figure}

\begin{figure}[t!]
    \centering
    \includegraphics[width=\textwidth]{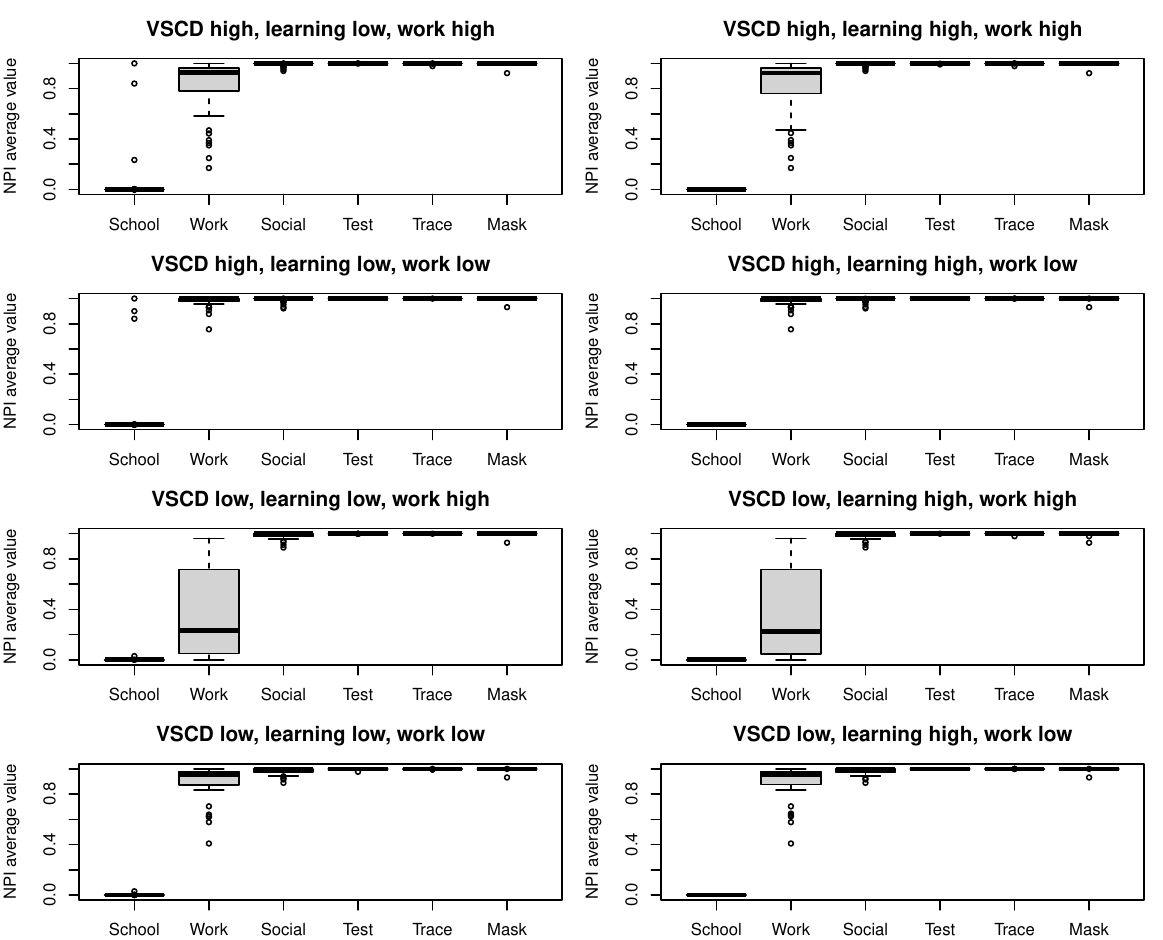}
    \caption{Sensitivity analysis for the optimal control results under regression model (ii). Boxplots of the average value of the optimal NPI policy over the year in each state.}
    \label{fig:sensitivity}
\end{figure}

\begin{figure}[t!]
    \centering
    \includegraphics[width=\textwidth]{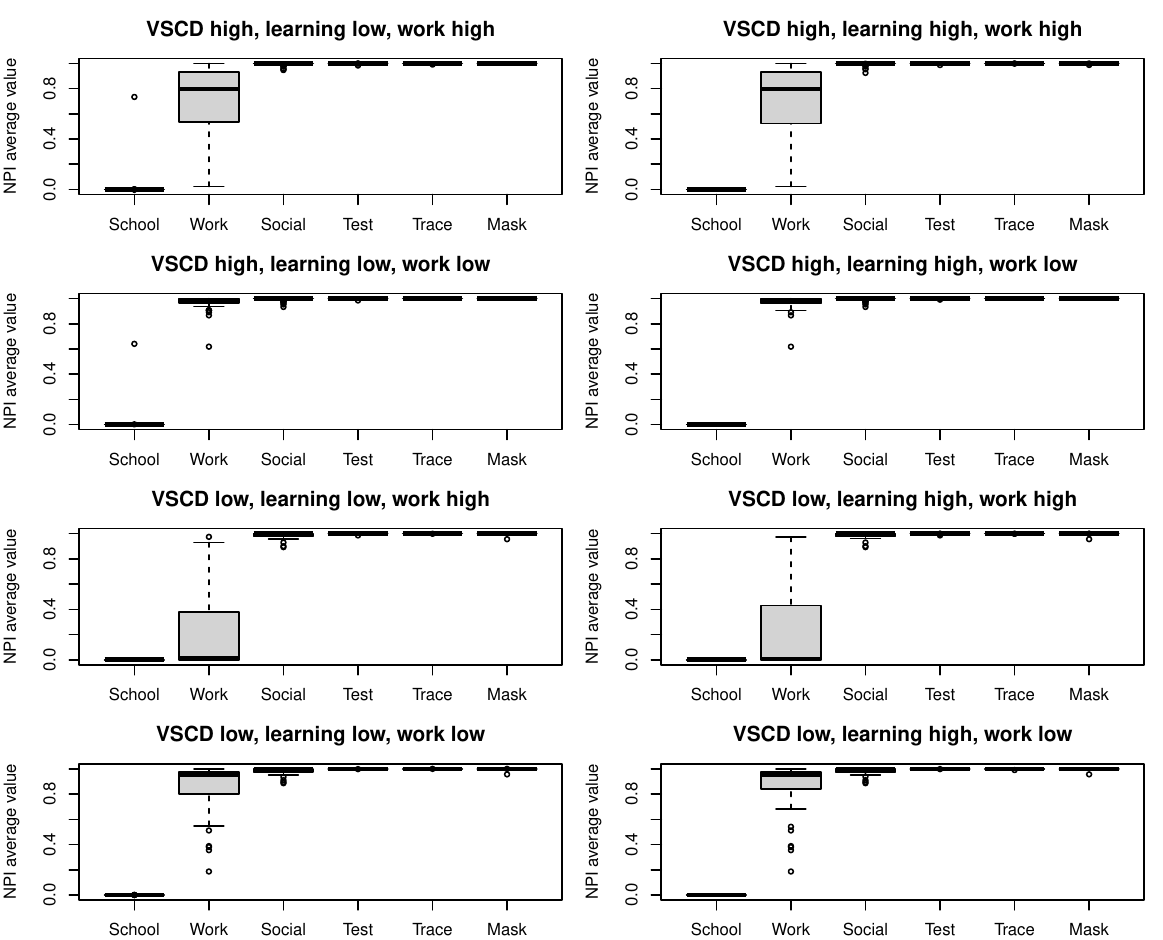}
    \caption{Sensitivity analysis for the optimal control results under regression model (iii). Boxplots of the average value of the optimal NPI policy over the year in each state.}
    \label{fig:sensitivity3}
\end{figure}

\begin{figure}[t!]
    \centering
    \includegraphics[width=\textwidth]{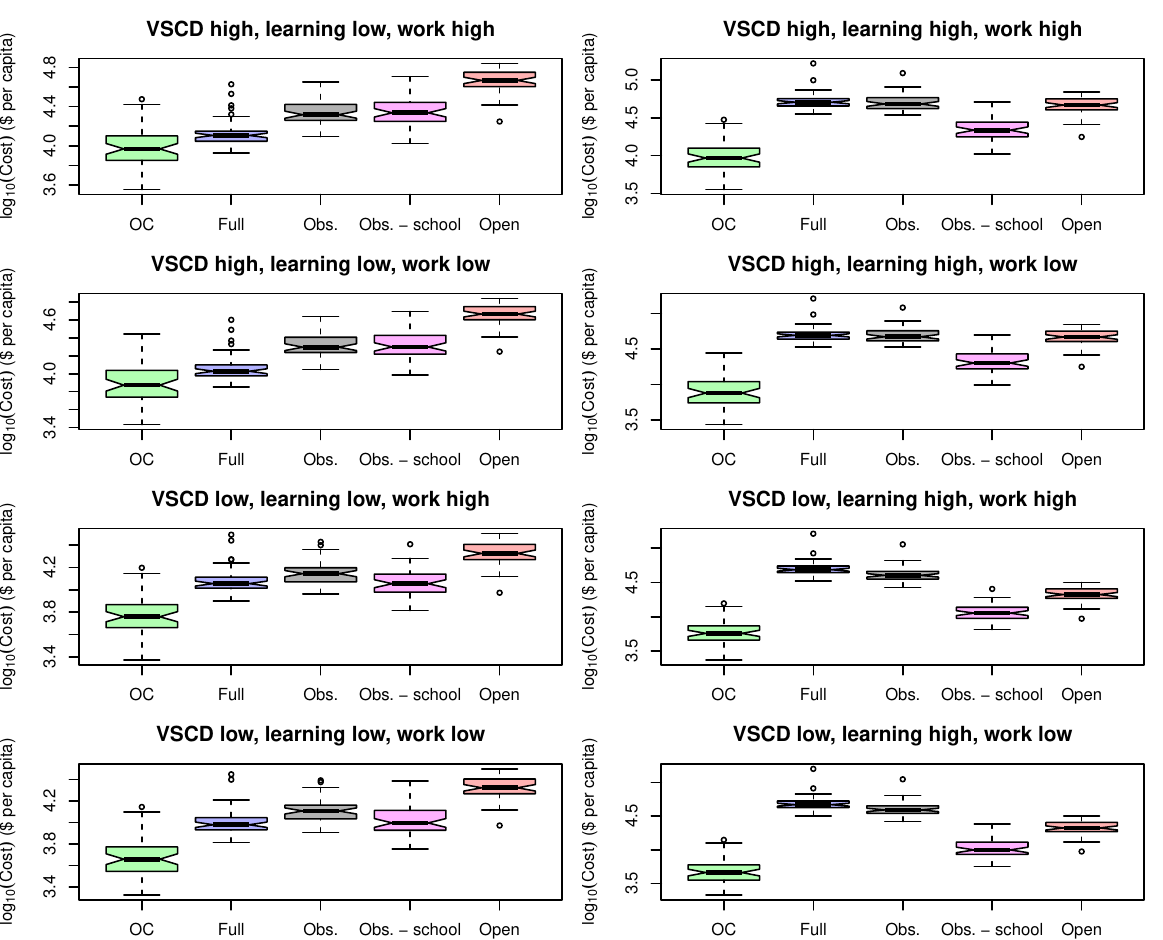}
    \caption{Sensitivity analysis for the costs of various policies under regression model (i). Boxplots of the log-scale total cost in USD2020 per capita incurred by the optimal control (OC), full lockdown (Full), observed (Obs.), observed minus school closures (Obs. - school), and fully open (Open) policies across states.}
    \label{fig:sensitivity-cost1}
\end{figure}

\begin{figure}[t!]
    \centering
    \includegraphics[width=\textwidth]{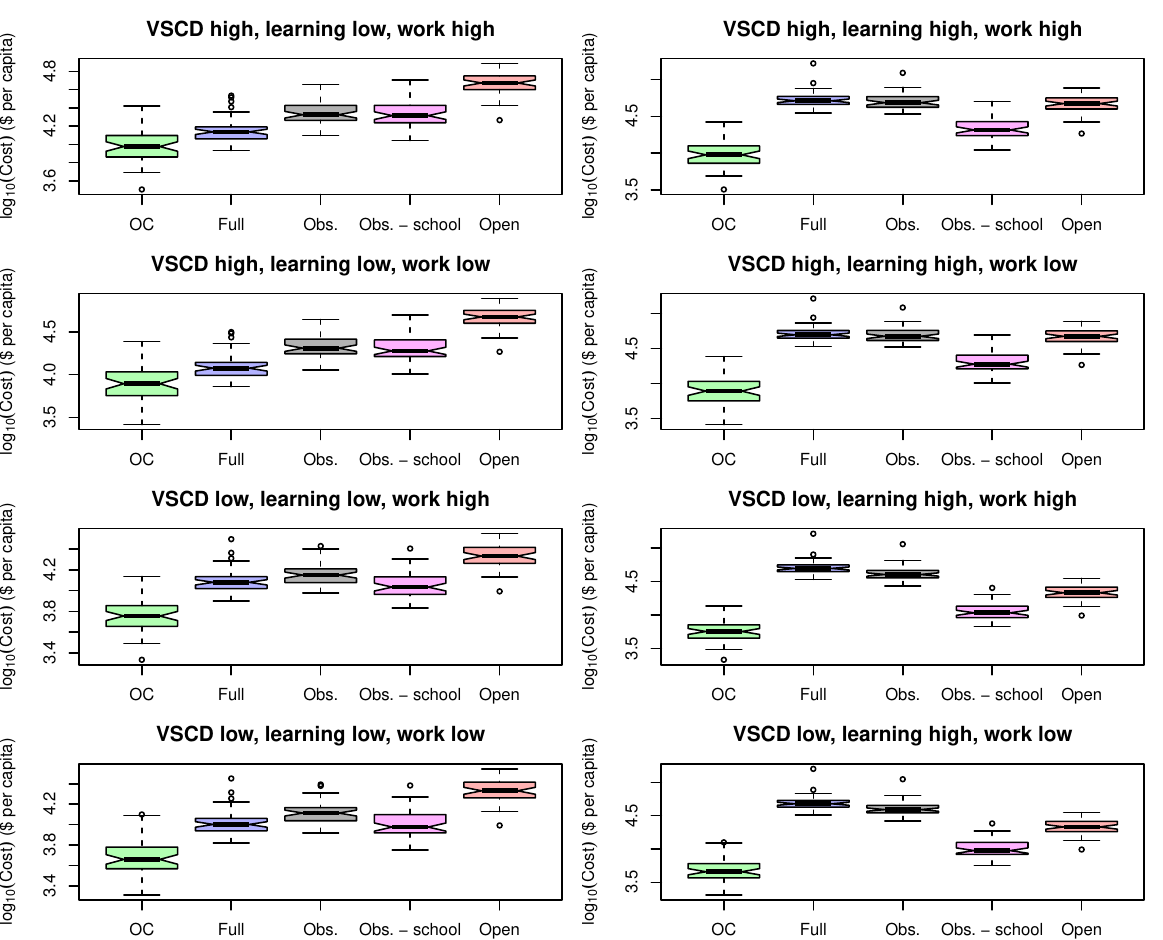}
    \caption{Sensitivity analysis for the costs of various policies under regression model (ii). Boxplots of the log-scale total cost in USD2020 per capita incurred by the optimal control (OC), full lockdown (Full), observed (Obs.), observed minus school closures (Obs. - school), and fully open (Open) policies across states.}
    \label{fig:sensitivity-cost}
\end{figure}

\begin{figure}[t!]
    \centering
    \includegraphics[width=\textwidth]{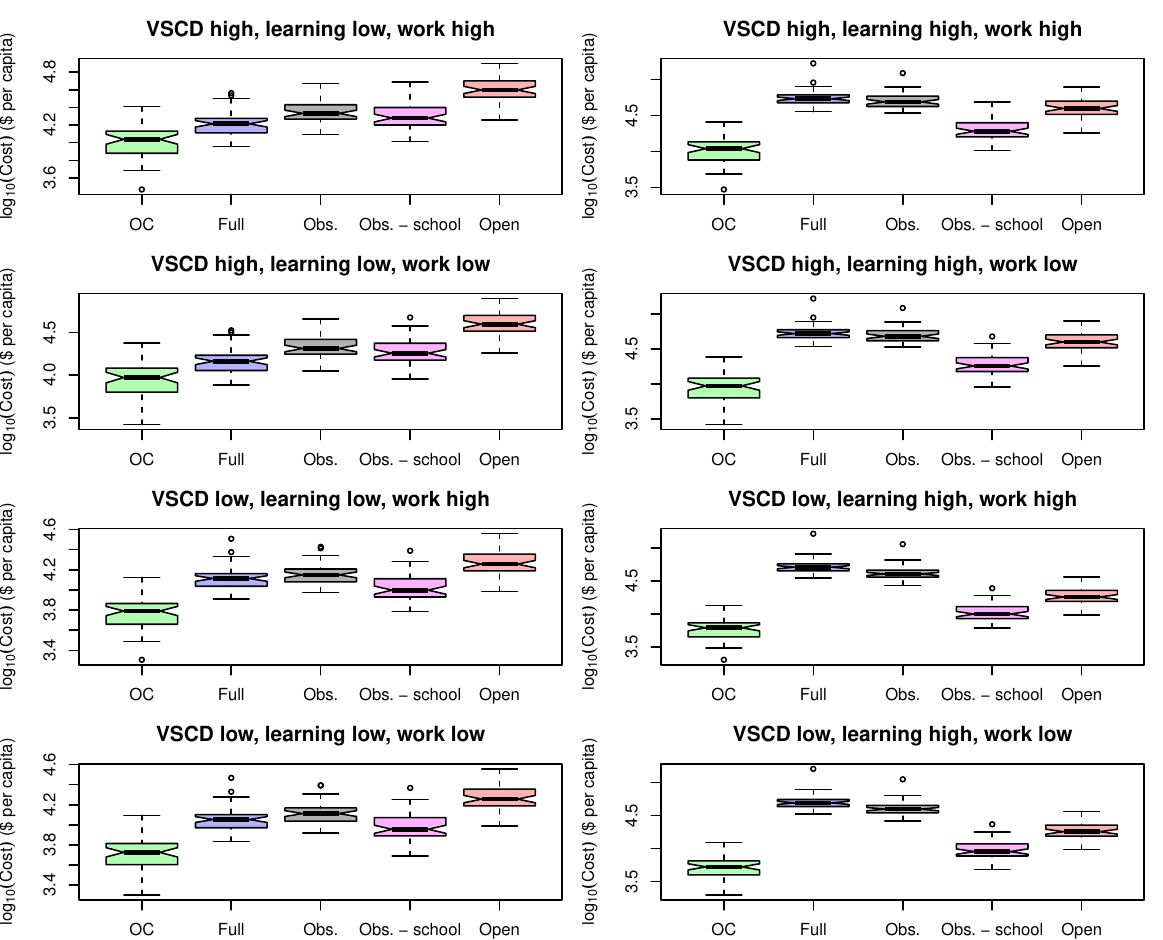}
    \caption{Sensitivity analysis for the costs of various policies under regression model (iii). Boxplots of the log-scale total cost in USD2020 per capita incurred by the optimal control (OC), full lockdown (Full), observed (Obs.), observed minus school closures (Obs. - school), and fully open (Open) policies across states.}
    \label{fig:sensitivity-cost3}
\end{figure}

Figures \ref{fig:sensitivity1}--\ref{fig:sensitivity-cost3} display the results of our sensitivity analysis across regression models (i)--(iii) and cost function specifications. 
Overall, our qualitative findings about the structure of optimal policies are robust across scenarios, with the main quantitative distinction being the optimal strength of workplace closure. We also find that the relative ranking by cost of the policies considered can vary across cost function specifications.

We vary the value of a statistical COVID death (VSCD), the cost of learning loss, and the cost of workplace closures across plausible ranges, which are given in Table \ref{tab:econ-params}. As noted above in Section \ref{sec:cost-social}, we do not vary the cost of social distancing measures as we are primarily interested in assessing the robustness of the optimal strategy and the relative costs of various policies rather than variation in the total cost incurred by each policy. We also omit sensitivity analysis for the costs of testing, tracing, and masking as they are highly cost-effective interventions under any reasonable variation in their costs. Indeed, these measures are all at least somewhat effective in reducing transmission and extremely cheap compared to infections, school and workplace closures, and social distancing mandates. 

Figures \ref{fig:sensitivity1}--\ref{fig:sensitivity3} exhibit boxplots of the average value of each NPI in the optimal strategy across states for models (i)--(iii), respectively.
We see that, barring a few outliers in some scenarios, school closures are never implemented in the optimal policy. This is the case even when we assume both a high VSCD equal to the VSL (\$10.63 million USD2020)---which does not adjust the VSCD for the age profile of COVID mortality---and a low cost of learning loss (9\% of GDP per 0.33 years)---which assumes that distance learning is 90\% as effective as in-person schooling---which are shown in the upper left panels. 
Across scenarios, workplace closures are implemented fairly consistently, but their optimal duration decreases as the cost of workplace closure increases or the VSCD decreases. 
In particular, in Figure \ref{fig:sensitivity3} (model (iii)), we see that the median optimal strength of workplace closure across states is 0 when we use a low VSCD and high cost of workplace closure. 
These conclusions are comparable to the results of \citet{barrot-business-effect}, who find that the cost-effectiveness of business closures (i.e., whether they produce a net benefit or loss) is sensitive to modeling assumptions and particularly the assumed value of a life-year.

Figures \ref{fig:sensitivity-cost1}--\ref{fig:sensitivity-cost3} show boxplots of the costs of various policies across states. Notably, when the cost of learning loss is high (right column), the total cost of the observed policy (i.e., the one that was actually implemented) minus school closures (denoted Obs. - school) is substantially lower than the observed (Obs.) and full lockdown (Full) policies. If we further assume that the VSCD is low (bottom right panels), Full exceeds the cost of Obs., which exceeds the cost of the fully open policy (Open). If both VSCD and the cost of learning loss are low (bottom left panels), Obs. - school becomes cheaper than Obs. and on par with Full. 
These patterns are consistent across models (i)--(iii).

\section{Incremental cost-effectiveness ratios (ICERs) in infectious disease}
\label{sec:results-icer}

\begin{figure}[t!]
    \centering
    \includegraphics[width=\textwidth]{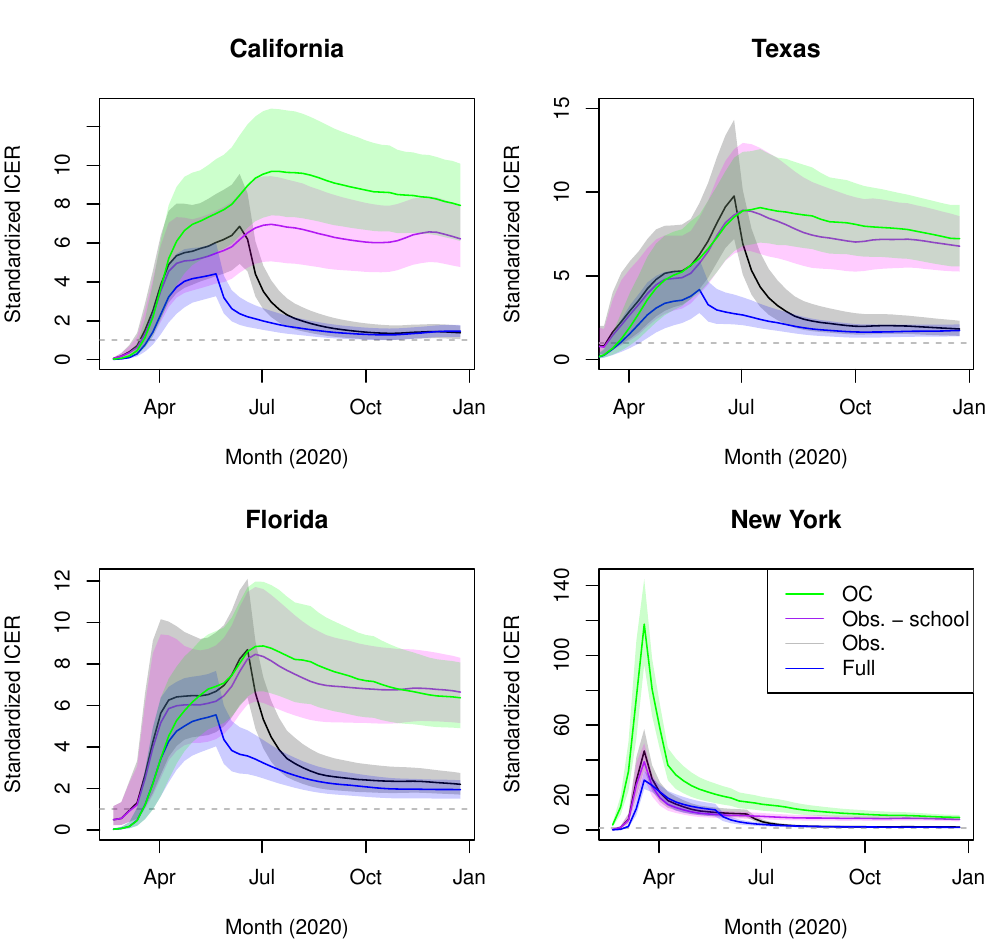}
    \caption{Posterior median and 50\% credible intervals for the cumulative standardized ICER of various policies in the four most populous states.}
    \label{fig:icer}
\end{figure}

\begin{figure}[t!]
    \centering
    \includegraphics[width=\textwidth]{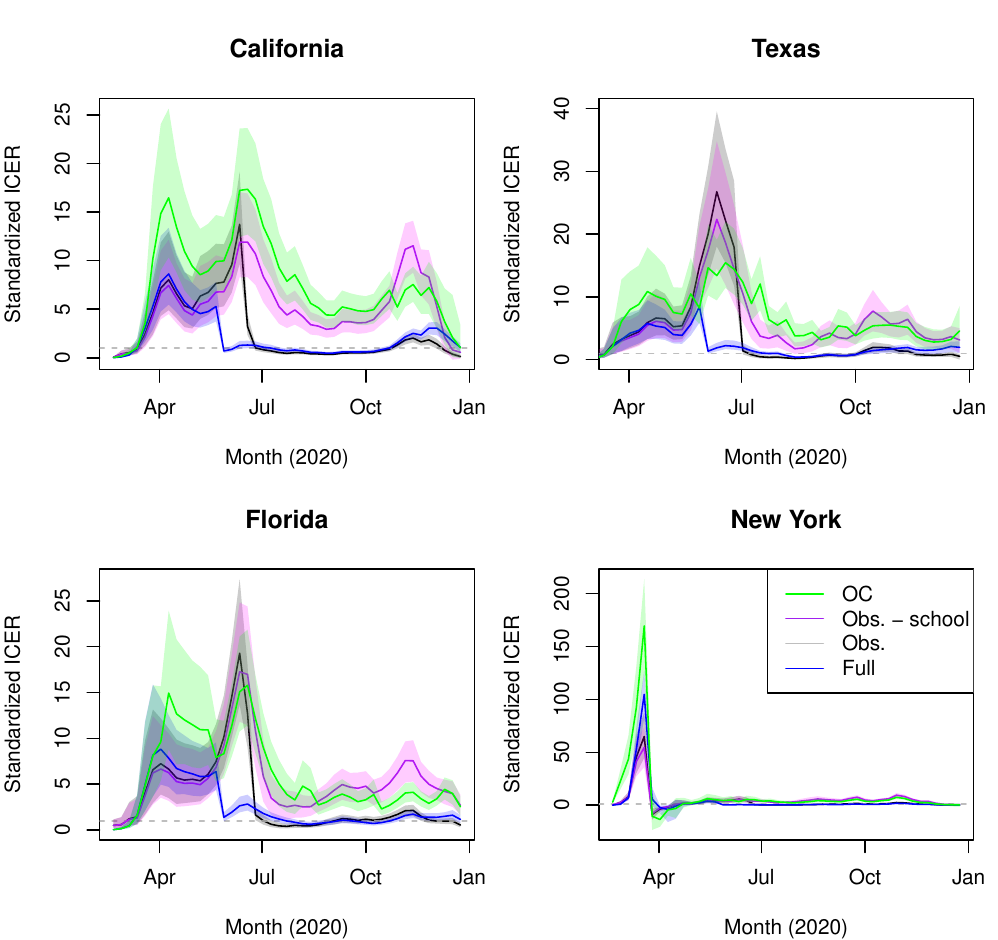}
    \caption{Posterior median and 50\% credible intervals for the weekly standardized ICER of various policies in the four most populous states.}
    \label{fig:icer-weekly}
\end{figure}

The incremental cost-effectiveness ratio (ICER) is a quantity widely used in the economic evaluation of health interventions. The ICER is defined as the monetary cost of an intervention divided by the benefit it produces (as measured by some target outcome) relative to a baseline, with a larger ICER often (mis)interpreted to mean that an intervention is less cost-effective \citep{paulden-icer}. In public health and infectious disease, outcomes of interest include the number of quality-adjusted life years (QALYs), disability-adjusted life-years (DALYs), infections, hospitalizations, or deaths averted by the intervention. Despite their ubiquity in the health economics literature, ICERs can be difficult to interpret and are often defined, calculated, and reported inconsistently, which severely limits their practical value in cost-effectiveness analysis \citep{paulden-icer,weinstein-icer}. As a result, ICER estimates for the same strategy can vary by orders of magnitude across studies and rankings of interventions based on their reported ICERs can yield counter-intuitive results, as evidenced by a recent systematic review of economic evaluations of COVID-19 interventions \citep{podolsky-review}. For example, \citet{podolsky-review} find that the median ICER of school closure across studies in their review is exceeded by that of vaccination, testing, facial covering, and stay-at-home policies.\footnote{This may be due to the fact that many cost-effectiveness analyses considering school closure fail to account for costs associated to student learning loss, as we discuss in Section \ref{sec:intro}.
} 
Furthermore, the median ICER of mask mandates exceeds that of school closure by an order of magnitude, and the median ICER of stay-at-home orders exceeds that of school closure by nearly 3 orders of magnitude. To address these issues, some have argued instead for the use of a policy's net benefit in decision-making \citep{paulden-icer,craig-net-benefit,stinnett-net-benefit}. We take a similar approach in Section \ref{sec:results-baseline}, reporting the expected total cost of various strategies, which facilitates straightforward comparison of their cost-effectiveness. 
Nevertheless, our methodology can shed light on the appropriate use of ICERs in the economic evaluation of interventions targeting infectious disease transmission.

We argue that ICERs in the context of infectious disease interventions should be: data-driven and account for uncertainty; not based solely on (poorly calibrated, overly simplistic, or deterministic) simulation models; reported based on well-defined interventions\footnote{Relevant details include the duration and strictness of implementation and the strength of adherence to the intervention.}; and calculated using the intervention's (causal) effect on the effective reproduction number $R_e(t)$, the relevant quantity governing infectious disease transmission, which is interpreted as the expected number of secondary cases resulting from an infection and can be estimated from clinical data in real time.
Related to this last point, we note that, for a given intervention, the ICER is a time- and context-specific quantity.\footnote{This is an important concern when working with ICERs. Outcomes, such as infections averted by an intervention, depend also on the implementation of other interventions (their timing, strength, duration) as well as other factors affecting disease transmission (e.g., the duration of the pandemic, the baseline $R_0$ value, other epidemiological parameters, voluntary social distancing and protective measures, exogenous shocks, new variants, etc.).} Therefore, in calculating and reporting the ICER, our results should account for or be as invariant as possible to contextual factors. We echo \citet{prager-flu-cost}, who highlight the ```...importance of including a broader set of causal factors to achieve more accurate estimates of the total economic impacts of not just pandemic influenza but biothreats in general.''

If we consider a blanket intervention that affects the population at large (e.g., social distancing measures), the intervention's reduction in $R_e(t)$ multiplied by the estimated size of the infectious population gives the number of infections averted by the intervention on day $t$, which forms the denominator of the ICER. On the other hand, if we consider a targeted intervention (e.g., case isolation), the number of infections averted can be estimated as the reduction in $R_e(t)$ multiplied by the number of treated subjects. Finally, the numerator of the ICER is the cost of implementing the intervention on day $t$.

To demonstrate, Figures \ref{fig:icer} and \ref{fig:icer-weekly} exhibit cumulative and weekly standardized ICERs of various strategies relative to the open policy (Open)---in which no NPIs are used---over time in the four most populous US states. We calculate these standardized ICERs as follows. For an NPI policy $u$ implemented on a specific day $t$, let
\begin{linenomath*}
\[
R_e(u,t) := R_0(u,t)S(t)
\]
\end{linenomath*}
denote the effective reproduction number on day $t$ under policy $u$, where (in an abuse of notation), $R_0(u,t)$ is the basic reproduction number defined by our NPI regression model \eqref{eq:rt-npi} and $S(t)$ is the susceptible fraction of the population. The number of infections averted by the policy $u$ on day $t$ relative to the open strategy is then
\begin{linenomath*}
\[
\nu_a(t) := N I(t)(R_e(\mathbf{0},t)-R_e(u,t)),
\]
\end{linenomath*}
where $N$ is the population size, $I(t)$ is the infectious fraction of the population, and $R_e(\mathbf{0},t)$ denotes the effective reproduction number under no interventions. In a given period of days $[T_1,T_2]$, the ICER of a policy $\mathbf{u}=\{u(t):t\in[T_1,T_2]\}$ is then
\begin{linenomath*}
\[
\text{ICER}_\mathbf{u}(T_1,T_2) = \frac{c_{\text{NPI}}(\mathbf{u})\cdot N}{\sum_{t=T_1}^{T_2} \nu_a(t)},
\]
\end{linenomath*}
where $c_{\text{NPI}}(\mathbf{u})$ is the per capita cost of the strategy, as defined in Section \ref{sec:methods-cost}. Finally, to define the standardized ICER plotted in Figures \ref{fig:icer} and \ref{fig:icer-weekly}, we convert infections to their monetary value using the cost of an infection $c_\nu$---defined in Section \ref{sec:cost-infection}---and take the reciprocal, such that the standardized ICER reports the ratio of the value of infections prevented to the cost of the intervention, which is a dimensionless quantity:
\begin{linenomath*}
\begin{equation}
\text{SICER}_\mathbf{u}(T_1,T_2) := \frac{\sum_{t=T_1}^{T_2} c_\nu\nu_a(t)}{c_{\text{NPI}}(\mathbf{u})\cdot N}.
\label{eq:sicer}
\end{equation}
\end{linenomath*}
Based on \eqref{eq:sicer}, in terms of net costs, a policy $\mathbf{u}$ relative to no intervention in time period $[T_1,T_2]$: breaks even if $\text{SICER}=1$; produces a net benefit if $\text{SICER}>1$; and produces a net loss if $\text{SICER}<1$. 

We now return to Figure \ref{fig:icer}, which displays the cumulative standardized ICER over time, $\{\text{SICER}_\mathbf{u}(1,t)\}_{t=1}^T$ for various policies $\mathbf{u}$. We see that, relative to Open, the different containment strategies are comparable in terms of cumulative cost-effectiveness early in the pandemic, with Obs. and Full becoming less cost-effective (but still producing a net benefit by the end of 2020), mainly due to the cost of school closures exceeding 16 weeks. The optimal control strategy OC is generally the most cost-effective by the end of the year, 
while Obs. - school is a close second. Note that, while OC and Obs. - school are nearly equally cost-effective by the end of 2020 in Florida, this does not imply that Obs. - school is also an optimal policy---this can only be determined by looking at the net benefit of each policy. Figure \ref{fig:icer-weekly} displays the standardized ICER of each policy in each week (conditional on that policy also being implemented in all weeks prior) relative to Open. Results are similar: OC and Obs. - school are similarly cost-effective and consistently more cost-effective than Obs. and Full; all containment strategies satisfy $\text{SICER}>1$ for most of the year, implying that they produce net benefits relative to no intervention. 

The optimal control strategy is determined by minimizing aggregate costs accrued over the year. We can solve this problem \emph{ex post}---after we have observed the pandemic play out---but, in principle, we cannot derive an optimal policy \emph{ex ante}, since we cannot see the future. Indeed, while SIR models have demonstrated remarkable utility in helping us understand infectious disease \citep{kermack-sir}, modeling studies carried out early in the pandemic (e.g., \citet{ferguson-npi-impacts}) generally failed to predict the complex and stochastic dynamics of SARS-CoV-2 (e.g., multiple waves, super-spreader events) depicted in Figure \ref{fig:us}. As we note in Section \ref{sec:results-fit}, even when we incorporate the effects of interventions into the model, we fail to explain a substantial portion of the temporal variation in transmission rates. At its face, this may seem like a disheartening realization. However, the trends in Figures \ref{fig:icer} and \ref{fig:icer-weekly} have important and encouraging implications for decision-making during pandemics. Specifically, they show that the OC policy---a relatively simple combination of testing, tracing, masking, social distancing, and reactive workplace closure---was consistently highly cost-effective on a weekly basis throughout the first year of the pandemic. This implies that, had we reasonable ballpark estimates of the costs and effects of interventions to work with early on, we could have determined a nearly-optimal strategy by choosing, at each point in time, the policy that greedily minimized the cost incurred in the next time step. In our context, we could not predict long-run transmission rates and therefore we did not need to; the myopic strategy would have produced a nearly-globally-optimal solution. A similar phenomenon has been documented in the control theory community \citep{recht-mpc}: the performance of control algorithms can be highly sensitive to modeling errors. Hence, if a model is misspecified, or if we can only poorly understand the behavior and evolution of a control system, simple algorithms tend to be more robust.

We turn now to the cost-effectiveness of school closure in particular. In Section \ref{sec:results-baseline}, we estimate that, relative to Obs. - school, the observed policy saved 
77,168 (12,268--235,954)
lives with the cost of school closures amounting to \$2 trillion, yielding an ICER of 
\$25.9 (8.4--156.4) million
per death prevented. 
By comparison, in a systematic review \citet{juneau-review} find that school closures during the 2009 H1N1 influenza pandemic cost \$9.86 million per death prevented, which implies that the cost of preventing a death due to H1N1 is on par with the VSL---taken to be \$10.63 million, in line with \citet{robinson-vsl}, in our study. 
In contrast, in our sensitivity analysis in Section \ref{sec:sensitivity}, we find that optimal policies for COVID involve no school closure beyond the usual 16 weeks of break per year even when the VSCD is assumed equal to the VSL and the cost of learning loss remains at the low value used in our baseline scenario. 

Qualitatively, our results are consistent with the findings of \citet{juneau-review} in other ways. Specifically, they find that: testing, tracing, and masking are among the most cost-effective measures; workplace and school closures are effective but costly, and hence the least cost-effective interventions; combinations of NPIs are more cost-effective than single interventions; and NPIs are more cost-effective when implemented early. While \citet{juneau-review} conclude that school closure is among the least cost-effective interventions based on the ICER, the value of this finding is limited for a number of reasons. 

Firstly, as \citet{paulden-icer} notes, it is difficult to decide based on the ICER alone if an intervention is cost-effective in a given setting, i.e., if it would be implemented in an optimal policy in conjunction with other interventions and other external factors. Indeed, as noted above, the cost-effectiveness of an NPI varies over time and across contexts, particularly as prevalence varies. 
(This is one factor explaining why earlier is better with regard to the timing of NPIs: as population immunity grows, NPI implementation yields diminishing returns.)
As such, calculating and ranking the ICERs of various interventions is not sufficient to determine which strategies are cost-effective \citep{paulden-icer}.
However, as we demonstrate in Section \ref{sec:results-baseline}, we can establish that school closure is not cost-effective by deriving the optimal policy, which frames the analysis in terms of the expected total cost---or, equivalently, the net benefit---of a policy. Our methodology goes beyond quantifying the cost-effectiveness ratios of different policies by determining which NPIs should have been used and when. 

Secondly, the results of \citet{juneau-review} are based on analysis of school closures during the 2009 H1N1 pandemic. The biology and epidemiology of the H1N1 flu differed from SARS-CoV-2 in important ways. In particular, it was evident early in the COVID-19 pandemic that SARS-CoV-2 was more virulent and more transmissible than H1N1, which left open the possibility that school closure would be cost-effective in combating SARS-CoV-2 transmission in 2020 despite its apparent lack of cost-effectiveness in 2009, as noted by \citet{pasquini-descomps-flu-cost-review,xue-school-cost}. 
Similarly, studying the cost-effectiveness of interventions in response to outbreaks of influenza, gastroenteritis, and chickenpox in France, \citet{adda-oc} finds that school closures are not cost-effective, but they ``...would become beneficial for epidemics characterized by a slightly more deadly strain.'' 
Indeed, \citet{dauelsberg-flu-schools,perlroth-flu-costs,milne-flu-cost,kelso-flu-cost,xue-school-cost} find that extended school closures are cost-effective for severe pandemics. We note that \citet{dauelsberg-flu-schools,perlroth-flu-costs,milne-flu-cost,kelso-flu-cost} do not account for costs associated to student learning loss and \citet{xue-school-cost} do not consider interventions other than school closure as available tools in their model.
As we show in Section \ref{sec:results-baseline}, in the context of COVID-19, school closures are not cost-effective.

\end{document}